\documentclass[a4paper,12pt]{article}
\usepackage[T1]{fontenc}
\usepackage{parskip}
\usepackage{times}
\usepackage{tikz}
\usepackage{rotating}
\usetikzlibrary{patterns}
\usepackage{kantlipsum}
\usepackage{pdfpages}
\usepackage{fancyhdr}
\usepackage{eurosym}
\usepackage{setspace}
\usepackage{verbatim,color}
\usepackage{amsmath,amssymb}
\usepackage{subcaption}
\usepackage{array}
\newcolumntype{H}{>{\setbox0=\hbox\bgroup}c<{\egroup}@{}}
\usepackage{float}
\usepackage{multirow,array}
\usepackage{graphicx}
\usepackage{pdflscape}
\usepackage{afterpage}
\usepackage{widetable}
\usepackage[british]{babel}
\usepackage{natbib}
\usepackage[flushleft]{threeparttable}
\makeatletter
\renewcommand\@biblabel[1]{}
\makeatother
\makeatletter
\def\blthanks{\gdef\@thefnmark{}\@footnotetext}
\makeatother

\begin{document}
	
	\title{\textbf{Does agricultural subsidies foster Italian southern farms? A Spatial Quantile Regression Approach
	}}
	\author {Marusca De Castris \thanks{Department of Political Sciences, University of Roma Tre}\and Daniele Di Gennaro \footnotemark[1] \thanks{Corresponding Author: daniele.digennaro@uniroma3.it}}  
	\date{\vspace{-3ex}}
	\maketitle
	\begin{abstract}
During the last decades, public policies become a central pillar in supporting and stabilising agricultural sector. Since 1962, EU policy-makers developed the so-called Common Agricultural Policy (CAP) to ensure competitiveness and a common market organisation for agricultural products. In 2003, CAP was substantially reformed by the development of a single payment scheme not linked to production (decoupling), but focusing on income stabilization and the sustainability of agricultural sector.
Notwithstanding farmers are highly dependent to public support, literature on the role played by the CAP in fostering agricultural performances is still scarce and fragmented, while major efforts are devoted to analyse the relationship between farm technical efficiency and subsidies \citep{bojnec2013farm,zhu2012technical}. Actual CAP policies increases performance differentials between Northern Central EU countries and peripheral regions \citep{giannakis2015highly}. 
This paper aims to evaluate the effectiveness of CAP in stimulate performances by focusing on Italian lagged Regions. Moreover, agricultural sector is deeply rooted in place-based production processes. In this sense, economic analysis which omit the presence of spatial dependence produce biased estimates of the performances. Therefore, this paper, using data on subsidies and economic results of farms from the RICA dataset which is part of the Farm Accountancy Data Network (FADN), proposes a spatial Augmented Cobb-Douglas Production Function to evaluate the effects of subsidies on farm's performances. 
The major innovation in this paper is the implementation of a micro-founded quantile version of a spatial lag model \citep{kim2004two} to examine how the impact of the subsidies may vary across the conditional distribution of agricultural performances. Results show an increasing shape which switch from negative to positive at the median and becomes statistical significant for higher quantiles. Additionally, spatial autocorrelation parameter is positive and significant across all the conditional distribution, suggesting the presence of significant spatial spillovers in agricultural performances.\\ 
 
\end{abstract}

\section{Introduction}
During the last decades, public policies become a central pillar in supporting and stabilising agricultural sector. Since 1962, EU policy-makers developed the so-called Common Agricultural Policy (CAP) to ensure competitiveness and a common market organisation for agricultural products. CAP was conceived as a flexible political tool able to identify and correct market inefficiencies by promoting modernisation processes and the renewal of agricultural work force. However, a substantial revision of the CAP was required to deal with the scale enlargement and the land abandonment in less favoured areas caused by globalization of commodity market\footnote{Literature on globalization of agricultural sector is widespread and heterogeneous. Key factors caused by globalization include: a major dependence on international trade to satisfy internal demand \citep{fader2013spatial,porkka2013food} and urbanisation processes \citep{lucas2004life}.}.\\ 
In 1992, ''MacSharry'' reform reverses the traditional perspective in CAP by shifting from product to producer support. This approach poses a strong emphasis on liberalization of global market by removing distortion caused by the support on prices and quantities\footnote{For a critical analysis of ''MacSharry'' Reform see: \cite{daugbjerg2003policy,folmer2013common}.}. This reforming process was consolidated in 2003 with the development of a single payment scheme. The major novelties introduced in 2003 includes the decoupling of the incentives from the production of any particular product and an increasing focus on a land-based approach and the sustainability of agricultural sector. Furthermore, EU Commission offers freedom of choice in how implement CAP. This has produced a series of differentiated public incentives, including regionalised payments on the basis of land extension, farm-specific support (e.g. France and Italy where payments are based on historical farm production levels) or a combination of both (e.g. Sweden and Germany).\\ 
In this context, farmers become highly dependent to public support. Notwithstanding, the role played by the CAP in determining and fostering agricultural sector is not yet fully investigated. This point has a twofold relevance. On one hand, European policy-makers are deeply committed in sustaining agricultural sector (43 \% of total budget during the programming period 2007-2013), while total public support reached 32 \% of agricultural income on average in the EU. Moreover, EU Commission, through the CAP, recognizes and promotes agricultural production and, in overall, rural areas as potential instruments to foster employment and competitiveness \citep{zhu2012technical,olper2014cap}. \\
On the other hand, decoupling public support to production stabilizes agricultural sector from prices  volatility \citep{hennessy1998production}.This income-stabilizing attribute has a corresponding insurance effect, which may affect optimal decisions. In this sense, policy change was expected to induce efficient farms to exit unprofitable businesses leading to aggregate productivity gains for the sector and supporting behavioural changes related to farms specialization \citep{kazukauskas2010analysing,kazukauskas2014impact}.\\
Literature presents lack of evidences on the role played by public policies in fostering the production of added value. This paper aims to develop an empirical framework by using an augmented Cobb-Douglas Production function which directly consider the impact of the subsidies. This analysis introduces the presence of spatial autocorrelation and heterogeneity to evaluate the occurrence of externalities. To better understand the relationships between added value, different forms of capital (Human, Fixed, Land and Public Subsidies) and spatial proximity we implement a Conditional Spatial Autoregressive Quantile Model. In other words, we test not only how neighbouring units can directly, or indirectly, affect the formation of added value and, in overall, economic performances, but we analyse the differential impact across all the conditional distribution.
\section{Literature Review}
The impact of agricultural policies is mainly investigated by analysing the effects on farm technical efficiency. Theoretical results on the link between subsidy-efficiency are ambiguous. \cite{zhu2012technical}, focusing on the case of Netherlands, Germany and Sweden, find a negative impact of the subsidies on farm efficiency. The aforementioned authors argue the existence of an inverse relation between subsidies and efficiency (i.e. higher level of policies lead to lower farmer efficiency).\\
The negative impact of the subsidies on productivity may result from allocative (and technical) efficiency losses owing to distortions in the production structure and factor use, soft budget constraints and the funding to less productive enterprises \citep{rizov2013cap}. Policy may negatively affect farm productivity by distorting production structure of recipient farms, leading to allocative inefficiency. Indeed, farmers may start investing in subsidy-seeking activities that are relatively less productive \citep{alston2002incidence}. However, \cite{bojnec2013farm}, analysing Slovenian farms, show an increase in allocative efficiency and profitability on small farms, while negative effects on technical and economic efficiency. \\
\cite{giannakis2015highly} highlight how actual CAP policies increases performance differentials between Northern and Central EU countries and peripheral regions (Mediterranean, Eastern, Northern Scandinavian). In this sense, they observe how countries with an high share of utilized agricultural land in less favoured areas, such as in the Mediterranean, are 94 \% less likely to attain high economic performances\footnote{Additional critiques to the CAP includes the access of new EU state members \citep{gorton2009folly,wegener2011administering}, the lack of institutional adaptation to CAP reform \citep{dwyer2007european} and failures in guarantee and preserving biodiversity \citep{pe2014eu}.}. \cite{rodriguez2004between}, looking at Obj.1 Regions, find significant effects of the CAP limited to investments in education and human capital, which represents only one-eight of the total commitments.\\
To summarize, \cite{minviel2017effect} by implementing a meta-regression analysis argue that the ambiguous results on farm technical efficiency are highly conditioned by different factors, including the empirical case, the type of policy considered and the methodology used. In this sense, they observe how empirical findings seem to be inconclusive and dependent from the case-study.\\
Starting from this assumption, we try to analyse the role played by the CAP by focusing on the impact on the formation of value added. This operation, has required the implementation of an augmented Cobb-Douglas Function. Since the seminal work of \cite{griliches1964research} the estimation of the aggregated production function becomes a common pratice in agricultural economics (see \cite{martin2001productivity,fleischer2005does,herrendorf2015sectoral} intra alia).\\
\cite{rizov2013cap} introducing the presence of the subsidies in an APF, find evidences on the inefficiencies of public incentives in EU-15 countries between 1990 and 2008. Furthermore, after decoupling reform the effect of subsidies on productivity was more nuanced and in several countries become positive. However, our analysis has not been restricted to the evaluation of the direct impact of the policies on added value, but we focus on the occurrence of spillover effects by including spatial econometric techniques. To the best of our knowledge, the unique applications in agricultural economics of a spatial augmented production function are in \cite{bille2015spatial} and \cite{deca2017rica}. While \cite{bille2015spatial} suggest an approach based on a two-step procedure to deal with unobserved spatial heterogeneity to demonstrate whether or not the model parameters show to be spatially clustered, \cite{deca2017rica} propose a spatial production function to estimate direct and indirect effects of the CAP to Italian farms. Their empirical funding highlight a significant and positive direct impact of the policies in fostering economic performances, at least partially, counterbalanced by negative spillovers. \\
Notwithstanding our approach is similar to the one presented in \cite{deca2017rica}, this paper proposes some novelties. As previously remarked, this paper starts from the idea of an Agricultural Production Function \footnote{The Agricultural Production Function is conceptually borrowed from the literature on Knowledge Production Function (KPF). \cite{marrocu2011proximity} and \cite{de2013regional} modify the traditional KPF to study the impact of geographical proximity in explaining spillover effects. Their approaches allow to consider spatial dependence for geo-referenced data and can be easily applied to agricultural economics \citep{autant2012spatial}. However, while the role of geographical proximity in determining knowledge spillovers is strongly supported by the literature, the major challenge of this work is to find evidences on spatial spillovers in the primary sector.} to analyse the impact of public policies on economic performances of Italian Farms. To better understand the relevance of public support in fostering the formation of added value we implement a multi-step analysis. In the first stage, we investigate the impact of the different typologies of capital by developing different agricultural production function by using a linear model. In the second-step, we test for the presence of spatial dependence to design a correct ''spatial'' production function. This step is fundamental to evaluate the occurrence of spillover effects due to public subsidies. In the last step, we implement a spatial autoregressive quantile regression to understand how economic performances are influenced by different level of capitals. While spatial quantile regression techniques cover a growing interest in economics, in agricultural it can be still considered as an uncommon practice \citep{kostov2009spatial}.\\
\cite{deca2017rica}, analysing the case of Italian Farms, demonstrate the presence of negative spillovers in the time frame between 2008 and 2009. In this analysis, we look only to Southern regions to understand if the localization in one of the peripheral Italian regions can influence the sign and the extension of the estimates \footnote{See \cite{giannakis2015highly}}. In this way, given the highly dependence of the farms located in lagged regions to public subsidies, we are able to investigate the differential impact on economic performances of the different typologies of capital and, in overall, understand if actual policies are correctly ''targeted'' or a ''corrective'' reform is needed to take into account the ''real'' needs of the primary sector. 

\section{Data and Exploratory Data Analysis}
In 2008, agricultural system was deeply conditioned by the global macroeconomic crisis. During this period, EU-27 deals with a contraction on agricultural production, in real terms, and a deflationary trend on prices. Moreover, the growth of input prices, due to the volatility on both energy and fertilizer markets, produces a reduction on added value per worker and employment. Under this perspective, Italian case is of particular interest.\\
Indeed, Italian added value at factor cost increased by 2.4\%, while the share to the formation of the GDP is stable at the 2.3\%. However, the good economic performances are not sufficient to reduce the gap between agriculture and the other sectors\footnote{Agricultural added value at factor cost per worker unit is 24316 \euro, a value corresponding to the 44\% of the average of Italian economy \citep{inea2008}.}. Italy is characterized by structural problems conditioning Agricultural performances. These issues include the presence of systematic differences between North and South, the lack of young farmers (only 13,2 \% has less than 44 years) and a land abandonment on marginal areas, especially for high altitude zones. The gap between North and South appear clear in terms of added value per worker unit. Indeed, although Southern regions grow more than Northern ones (3.5\% vs 0.6\%), the average added value per worker unit is still well below Italian average (19300 vs 22000 \euro). Considering the weakness of Southern agriculture \footnote{The development gap between North and South is not limited to primary sector. Indeed, regions located in the Mezzogiorno are recognized, by European Commission, as less developed and transition regions. This classification is based on the levels of GDP and employment. Less developed (resp. transition) regions include the areas where GDP per head is less than 75\% (resp. between 75 and 90\%) of the EU average. This includes nearly all the regions of the new member states, most of Southern Italy, Greece and Portugal, and some parts of the United Kingdom and Spain. The Italian less developed regions are Campania, Apulia, Calabria, Sicily and Basilicata, while transition regions are Abruzzo, Molise and Sardinia. However, we exclude Campania from our analysis for a lack of comparability with the other southern regions, while Abruzzo and Molise are not considered for a lack of information about the farms included in our sample.} and its relationship between economic performances and public policies is the focus point in this paper.\\
By considering information extrapolated by RICA dataset\footnote{Rica is part of the European Farm Accountancy data network (FADN) and it represents the only harmonized survey to collect micro-economic data on firms operating in agricultural. Italian RICA collects information on 11000 farms sampled at regional level.RICA's field of observation considers only the farm with at least 1 hectare of UAA or a production value greater than 2500 Euros.} for 2008 we introduce in our analysis five different variables. \\

\begin{center}
INSERT TABLE \ref{tab:var}
\end{center}
\pagebreak
Table \ref{tab:var} considers all the major determinants on value added formation: Labour, Fixed Capital, Land and Policies. By applying differentiated formulation of the Cobb-Douglas APF we can understand the marginal impact of all the different variables in determining and stimulating value added. Subsidies are considered as a global indicator of public expenses to stimulate and sustain agricultural activities. In this sense, the total amount of subsidies is obtained by adding all the amount of the different public instruments allocated to every farm, independently from their source or objective (i.e. we do not distinguish between National or European fund or between policies devoted to current activities, rural development or capital subsidies). \\
Notwithstanding this paper is focused on the impact of public policies in value added we do not consider a counterfactual approach. This choice is related to the peculiarity of the agricultural sector. Indeed, 90\% of the farms in RICA dataset are subsidised. However, most of the subsidies are decoupled by production objectives and, in many case, constitutes a sort of income maintenance. In this sense, we exclude from our analysis all the farms which are not subsidized and we consider the incremental impact of the policies (i.e. effect of receiving an additional euro of public funds). As previously remarked, we include in our analysis all the farms located in Apulia, Basilicata, Calabria, Sicily and Sardinia, which constitutes part of Italian lagged regions. In this way, we build a dataset composed by 1298 farms.\\

\begin{figure}[htbp!]
	\centering
	\caption{Spatial Distribution}
	\label{fig:sd}
	\begin{subfigure}{.5\textwidth}
		\centering
		\includegraphics[width=.8\linewidth,height=5cm]{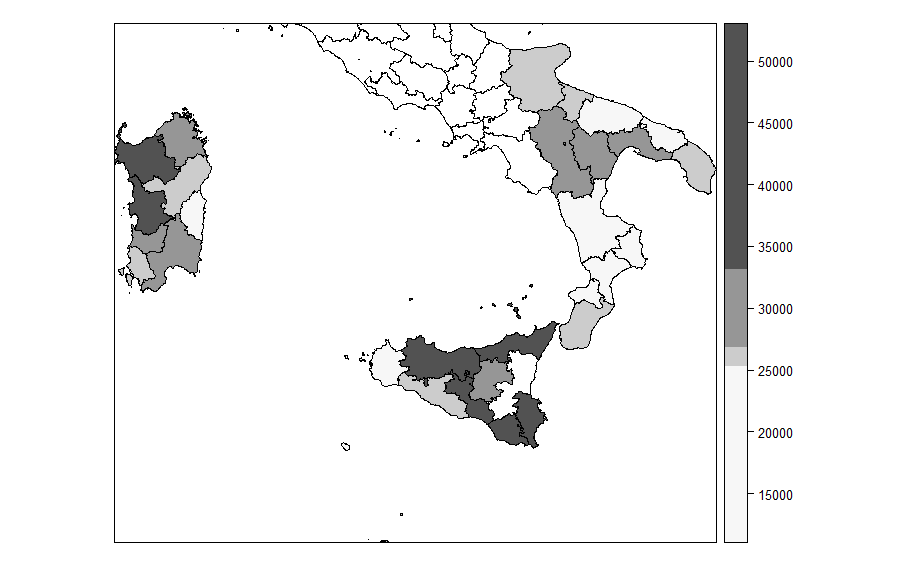}
		\caption{}
	\end{subfigure}%
	\begin{subfigure}{.5\textwidth}
		\centering
		\includegraphics[width=.8\linewidth,height=5cm]{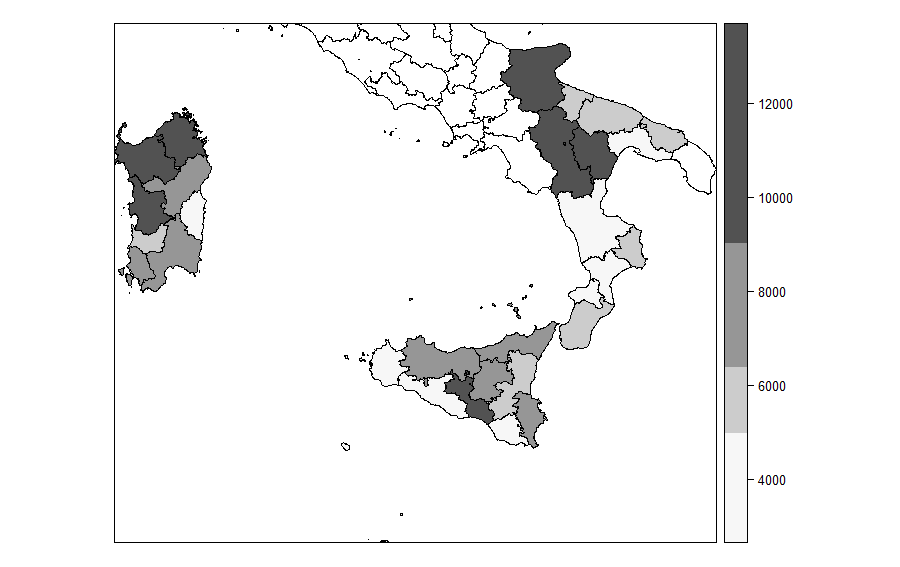}
		\caption{}
	\end{subfigure}
\caption*{Panel (a) (resp. b) shows spatial distribution of the Value Added per Labour Unit (resp. Subsidies per Labour Unit) aggregated at NUTS-III level.}  
\end{figure}

\begin{center}
	INSERT FIGURE \ref{fig:sd}
\end{center}
\vspace{13pt}

The first step of our exploratory spatial data analysis passes through the aggregation of the farms at NUTS-III level. Figure \ref{fig:sd} shows the spatial distribution of provincial added value and amount of subsidies. Both variables are standardized by unit of labour. In detail, in panel (a) we observe how value added exhibits a spatial polarization (i.e. similar level are concentrated in neighbouring areas), especially in Sicily and in the west of Sardinia.\\
On the other hand, panel (b), presenting spatial distribution of subsidies per labour unit, demonstrates the existence of a regional cross-border spatial pattern between Basilicata and the north of Apulia (i.e. the province of Foggia) and a spatial polarization of the subsidies in Sardinia. However, look at aggregated variables is not sufficient to individuate the presence of spatial clusters.\\
At this end, by considering observational data we estimate two different measure of spatial association: global and local Moran Index\footnote{Global spatial analysis or global spatial autocorrelation analysis yields only one statistic to summarize the whole study area. In other words, global analysis assumes homogeneity. If that assumption does not hold, then having only one statistic does not make sense as the statistic should differ over space. To allow for differentiated spatial patterns for each location, Local Moran I is a better option \cite{anselin1995local}.}. The spatial weight is computed by using a row-standardized matrix based on a cut-off distance\footnote{In the remainder of this paper, three different cut-off distances are considered to confer robustness to our results. The baseline cut-off is $d_{min}=33 km$, corresponding to the minimum distance for whom every farms have at least one neighbour. Alternative cut-off considered are $d_{1.25}=1.25*d_{min}=41 km$ and $d_{1.50}=1.50*d_{min}=50 km$.}.\\ In other words, two units are neighbours if their distance is less than d. This assumption can be easily resumed, in analytical form, as it follows:\\ 
\begin{equation*}
w_{ij}=
\begin{cases}
\frac{1}{\sum_{j=1}^{n}w_{ij}},\text{        } \text{        } \text{if    } 0\leq d_{ij} \leq d \\ \\
0, \text{        } \text{        } \text{        } \text{        } \text{        } \text{        }  \text{        } \text{        } \text{        } \text{        } \text{        } \text {if    } d_{ij}\geq d 
\end{cases}
\end{equation*}  

where $d$ corresponds to the different cut-off distance considered (i.e. 33 km, 41 km and 50 km).

\begin{figure}[htbp!]
	\centering
	\caption{Local Moran Index}
	\label{fig:locmor}
	\begin{subfigure}{.5\textwidth}
		\centering
		\includegraphics[width=.8\linewidth,height=5cm]{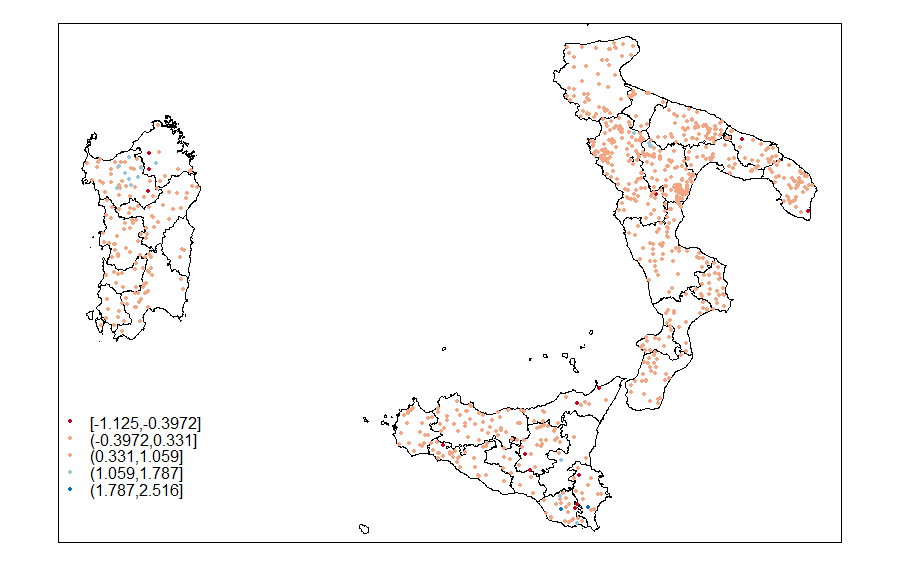}
		\caption{}
	\end{subfigure}%
	\begin{subfigure}{.5\textwidth}
		\centering
		\includegraphics[width=.8\linewidth,height=5cm]{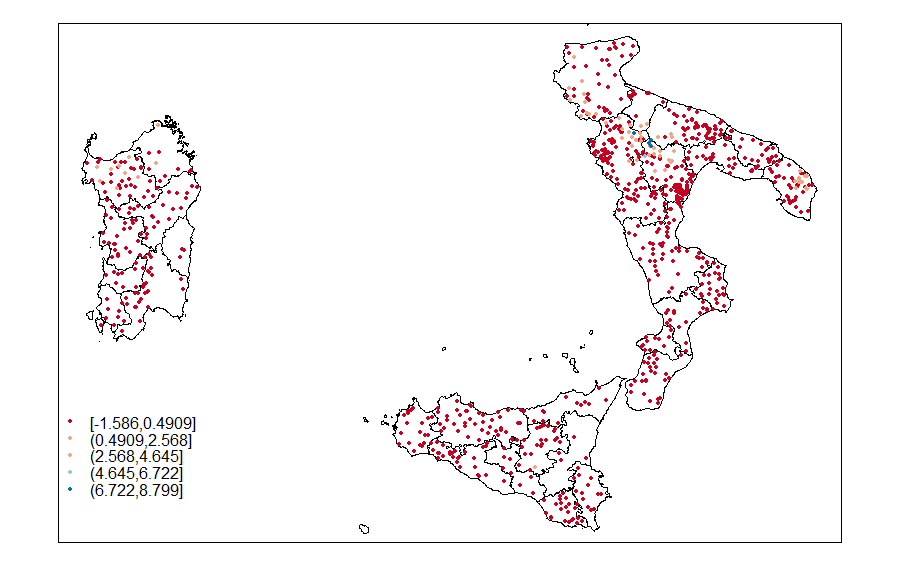}
		\caption{}
	\end{subfigure}
	\caption*{Panel (a) (resp. b) shows the distribution of the Local Moran Index of the Value Added (resp. Subsidies). Both variables are normalized by Unit of Labour. Global Moran Index for Added Value (resp. Subsidies) is equal to 0.053 (resp. 0.156) with a p-value of 0 (resp. 0). In this figure, we report only results on a cut-off equal to 33 km. However, results for other distances are identical.}  
\end{figure}

\begin{figure}[htbp!]
	\centering
	\caption{Local Clusters}
	\label{fig:locclust}
	\begin{subfigure}{.5\textwidth}
		\centering
		\includegraphics[width=.8\linewidth,height=5cm]{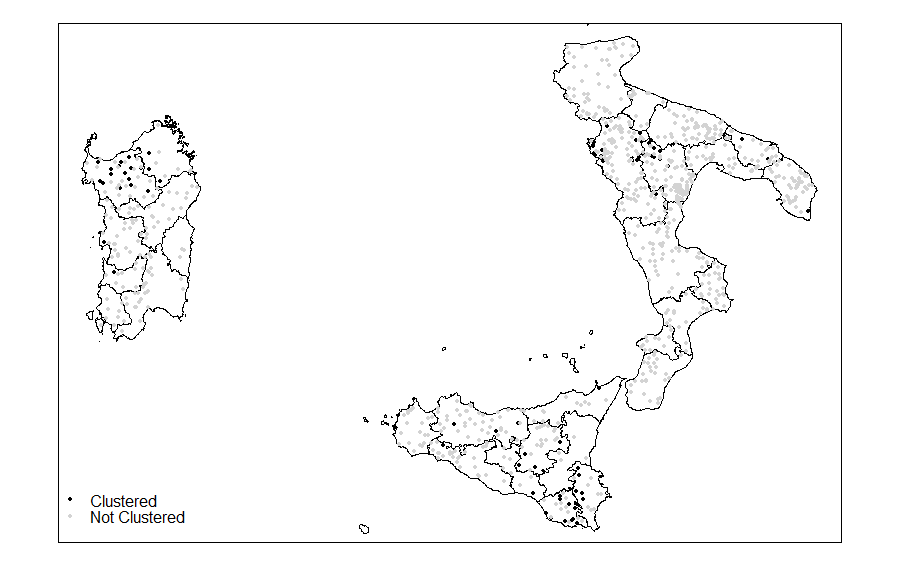}
		\caption{}
	\end{subfigure}%
	\begin{subfigure}{.5\textwidth}
		\centering
		\includegraphics[width=.8\linewidth,height=5cm]{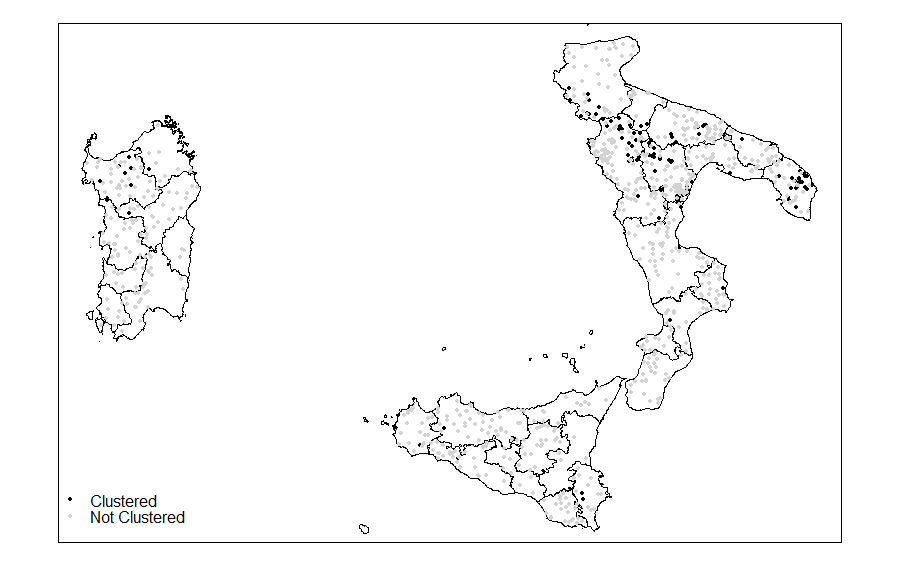}
		\caption{}
	\end{subfigure}
	\caption*{Panel (a) (resp. b) shows the distribution of the Local Moran Index of the Value Added (resp. Subsidies). Both variables are normalized by Unit of Labour. Units considered as clustered are the ones who present a p-value lower than 0.05. In this figure, we report only results on a cut-off equal to 33 km. However, results for other distances are identical.}  
\end{figure}

\begin{center}
	INSERT FIGURES \ref{fig:locmor} AND \ref{fig:locclust}
\end{center}

Global Moran I for both added value per labour unit and subsidies per labour unit are significant, highlighting the presence of spatial autocorrelation. Going in depth, figure \ref{fig:locmor}, relaxing the assumption of homogeneous spatial autocorrelation across the space, shows the distribution of local Moran index. The spatial distribution of the local Moran I confirms the major findings in figure \ref{fig:sd}. For sake of clarity,Figure \ref{fig:locclust} highlights the presence of local clusters \footnote{Local cluster are identified by isolating unit presenting a p-value lower than 0.05 .}. Panel (a), indicating added value per labour unit, shows the presence of isolated clusters in Sicily and north of the Sardinia (high-high cluster) and in part of Basilicata and Apulia (low-low cluster). Conversely, panel (b) presents evidences of high-high clusters between the province of Foggia and Basilicata and in the northern part of the Sardinia, while significant low-low clusters characterized the eastern part of the Apulia (i.e. Lecce). To resume, local and, in particular, global Moran indexes suggest the presence of a non-random spatial distribution for both value added and, in particular, subsidies. In this sense, in the remainder of this paper we check for the inclusion of the spatial dimension in evaluating the impact of the different components in stimulating value added. \\

\section{Methods and Results}

The empirical analysis in this paper relies on an unrestricted Cobb-Douglas APF .The traditional approach in estimating Cobb-Douglas APF is based on the estimation of the relationship between the value of production and technological input. However, in this work we aim to find evidences on the impact of the subsidies on economic performances of the farm. In this sense, we do not look at production value by itself, but, by using value added, we consider the relative impact of the different input in fostering agricultural performances. The baseline model assumes the form in:
\vspace{-1ex}
\begin{equation}
Y=AL^{\alpha}K^{\beta}G^{\gamma}S^{\delta}
\end{equation}
Where Y is the Added Value, A the Total Factor Productivity, L the labour units, K the fixed capital, G represents the ground extension (i.e. UAA) and S the total amount of subsidies. Following the traditional approach, we implement a log transformation of the APF. The final model assumes the following formulation:
\vspace{-1ex}
\begin{equation}
ln(Y)=ln A+\alpha* ln L+\beta* ln K+ \gamma*ln G+\delta* ln S
\end{equation}
Under this formulation we are able to identify and estimate all the parameters. Applying a log transformation of the APF is a completely risk-free assumption, even if it implies a different approach in discussing parameter estimates. Indeed, parameter estimate represents the elasticity (i.e. marginal impact in the output on a percentage change of the input).\\
The empirical strategy implemented in this paper is based on a multi-stage analysis. In the first step, we look at the effect of the different specification of the Cobb-Douglas production function by estimating an OLS. In this way, we can obtain baseline results of our approach by omitting the presence of spatial dependence. Furthermore, OLS estimates are used to evaluate if a spatial approach is preferable. On this regard, we follow the guidelines in \cite{anselin1996simple}. Aforementioned authors suggests to test OLS estimates with a Lagrange Multiplier Tests to identify the presence of spatial autocorrelation and/or spatial heterogeneity. In the last stage of this work, we implement a Spatial Autoregressive Quantile Regression to deepen how economic performances are influenced by different level of the covariates.  

\subsection{Linear Model}

Literature on the impact of public policies on agricultural performances is still scarce \citep{deca2017rica}. In this sense, to fully understand the impact of different inputs on the formation of the added value and, consequently, on performances we estimate a micro-founded model by using observational data. Furthermore, to observe the relative impact of all the variables we estimate 4 different specification of models. The baseline approach starts from the traditional Cobb-Douglas Function and considers two only inputs: labour and fixed capital, while the alternative specifications are modelled by including land and public expenses. 
\begin{center}
	INSERT TABLE \ref{tab:ols}
\end{center}
The estimates in Table \ref{tab:ols} show that the most effective input in improving performances is the labour. Indeed, L parameter passes from 0.918 to a minimum of 0.866 which means that a marginal increment of a percentage in labour is almost completely absorbed in terms of marginal increment of performances. While results on the marginal impact of the land are negligible, the estimates on both fixed capital and subsidies underline their relative importance on fostering performances. Indeed, a percentage change on level of both fixed and public capital has an identical impact by contributing to a 0.15 percent increase in performances. In overall, results are in line with the one presented in \cite{deca2017rica}.\\
Lagrange multiplier test does not allows to obtain a clear indication on the choice between a Spatial Autoregressive and a Spatial Error Model, while a so-called SARMA (Spatial AutoRegressive Moving Average) is preferable. This model combining both a spatial autoregressive and a spatial error component and implies a complex spatial error structure for error externalities. These results are consistent for all the different specification of the model and robust for the three cut-off distances considered.
\subsection{SARMA Estimates}
After estimating the OLS model which provides baseline results on the impact of subsidies in affecting performances, we include the presence of spatial heterogeneity. Given the meaningful evidence in favour of a SARMA model, we estimate a general ''spatial'' model of the form:
\vspace{-1ex}
\begin{align}
\label{eq:logsp}
\begin{split}
y_i &=a_i+\rho* (Wy)_i+\alpha* l_i+\beta* k_i+ \gamma*g_i+\delta*s_i+ u_i,
\\
u_i &=\lambda*Wu_i+\epsilon_i.
\end{split}
\end{align}
In Equation (\ref{eq:logsp}) lower case variables are indicated in terms of logarithm. \cite{anselin2003spatial} highlights how a SARMA model allows to combine local and global effects. Indeed, a SARMA is composed by a (local) spatial moving average component and a (global) spatial autoregressive process. SARMA required the estimation of two different spatial parameter: $\rho$ (AR component) and $\lambda$ (MA component). Endogeneity in response variable is considered by a multi step GM/IV estimation \citep{kelejan1998,kelejian2010specification,arraiz2010spatial}. As in linear model, to fully understand the differential impact of K,L,G and S we present the estimates of 4 different models obtained by combining the 4 typologies of capital. Moreover, different weight matrix are considered to provide robust estimates. 
\begin{center}
	INSERT TABLE \ref{tab:sarma}
\end{center}
Results in table \ref{tab:sarma} confirm how the labour is the factor which has a stronger impact in improving economic performances going from 0.89 (model 1) to 0.82 (full model). The extension and the significance of all the other parameter estimates are still in line with the results of the linear model. Interestingly, both $\rho$ and $\lambda$ are positive and significant confirming the hypothesis of a strong spatial dependence of our data. In overall, MA component overcomes the estimates of the AR parameter, indicating a greater ''local'' spatial impact.\\
Results are robust to the different weight matrix, even if spatial parameters offers some interesting intuition. Indeed, the distance between $\lambda$ and $\rho$ grow up passing from 33 to 41 km, while global effect becomes more important for higher distances. However, this parameter does not provide an estimate of the spillover effects. Indeed, in models containing spatial lags of the dependent variables, interpretation of the parameters becomes richer and more complicated. A number of researchers have noted that models containing spatial lags of the dependent variable require special interpretation of the parameters \citep{kim2003measuring,anselin2006interpolation}.\\
In estimating direct and indirect effects, a SARMA model does not differ from a Simple Spatial Autoregressive model. Indeed, \cite{lesage2009introduction} demonstrate how equation (\ref{eq:logsp}) can be rewrite as:
\vspace{-1ex}
\begin{equation}
\label{eq:sarma}
y_i =(I-\rho W)^{-1}(a_i+XB)+(I-\rho W)^{-1}*(I-\lambda W)^{-1}\epsilon_i
\end{equation}
where X and B represents, respectively, the vector of covariates and parameters. By taking the expectations of (\ref{eq:sarma}), we obtain the same DGP of a SAR:
\begin{equation}
\label{eq:sar}
y_i =(I-\rho W)^{-1}(a_i+XB)
\end{equation}
Therefore, SARMA models concentrates on a more elaborate model for the disturbances, whereas the interpretation of parameter estimates is identical to SAR process. On this issue, is widely recognized how the marginal effect of every variable in a spatial lag model is not the parameter by itself but becomes a composite function. By considering, in way of example, the derivative of y with respect to K we have :
\begin{equation}
\frac{dy}{dK}=(I-\rho W)^{-1}*\beta
\end{equation}
Clearly, the matrix $(I-\rho W)$ can be inverted only in the case in which $|\rho|<1$. However, by inverting the matrix we obtain a summary of the indirect effects by considering the average of the summation by row (or column) of the off-diagonal elements, while the direct effects are estimated by averaging the diagonal element of the inverse. To conclude, a summary of the total effect is estimated by averaging the summation by row (or column) of all the elements in the inverse matrix. The results of the spillover effects are in Table \ref{tab:marg}.
\begin{center}
	INSERT TABLE \ref{tab:marg}
\end{center}
Results highlights the presence of positive and significant indirect effects, independently from variables, models and spatial weight matrices considered. Also in this case, the indirect impact is of greater intensity for the human capital, while there are few evidences in favour of the occurrence of spillover effects for the land. As previously argued, fixed and public capitals are significant and of similar intensity.\\
In overall, direct effect dominate spillovers. However, considering the positivity of both direct and spillover effects, total impact is wider than the estimation in the SARMA. Furthermore, the role of the distance on the extension of spillover is evident, while direct effects are stable across different weight matrices. In detail, spillovers intensity does not change between 33 and 41 km, but for higher distance (50 km) becomes wider. This results are line with the evidences in Table \ref{tab:sarma}.\\
Considering the relevant impact of the different typologies of capitals and the occurrence of spillover effects, a further step in our analysis is required. In the remainder of this paper, we remove the hypothesis of homogeneous effects between different level of the variable by introducing a Spatial Quantile Regression.\\
\subsection{Spatial Quantile Regression}
\begin{figure}[ht]
	\centering
	\caption{Spatial Quantile Regression for Fixed Capital}
	\label{fig:sqr_k}
	\begin{subfigure}{.33\textwidth}
		\centering
		\includegraphics[width=.95\linewidth,height=5cm]{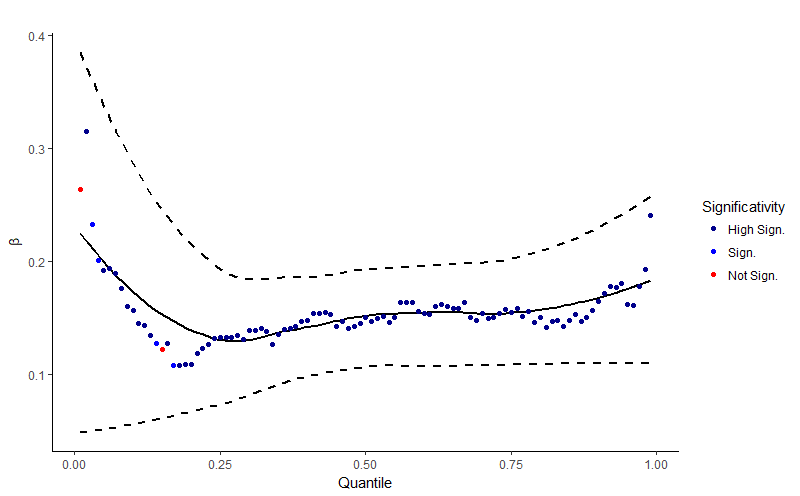}
		\caption{33 km}
	\end{subfigure}%
	\begin{subfigure}{.33\textwidth}
		\centering
		\includegraphics[width=.95\linewidth,height=5cm]{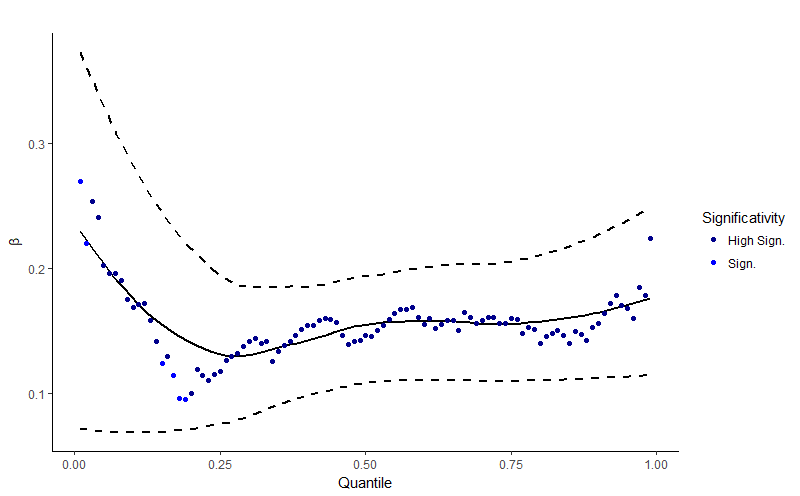}
		\caption{41 km}
	\end{subfigure}%
	\begin{subfigure}{.33\textwidth}
		\centering
		\includegraphics[width=.95\linewidth,height=5cm]{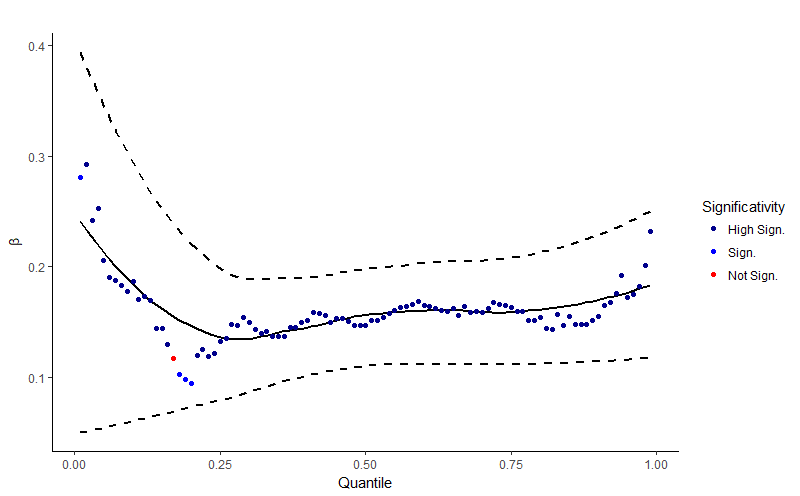}
		\caption{50 km}
	\end{subfigure}
	\caption*{\footnotesize {\textbf{Note:} Figure \ref{fig:sqr_k} shows the estimates of the Spatial AR model for every quantile. Solid line represents the smoothed function of the estimates, while dashed lines are the confidence interval at 95\%. Statistical significance is reported by different colours: Dark Blue= 0.01, Light Blue= 0.05, Red no significance.}}

\end{figure}

\begin{figure}[ht]
	\centering
	\caption{Spatial Quantile Regression for Labour}
	\label{fig:sqr_l}
	\begin{subfigure}{.33\textwidth}
		\centering
		\includegraphics[width=.95\linewidth,height=5cm]{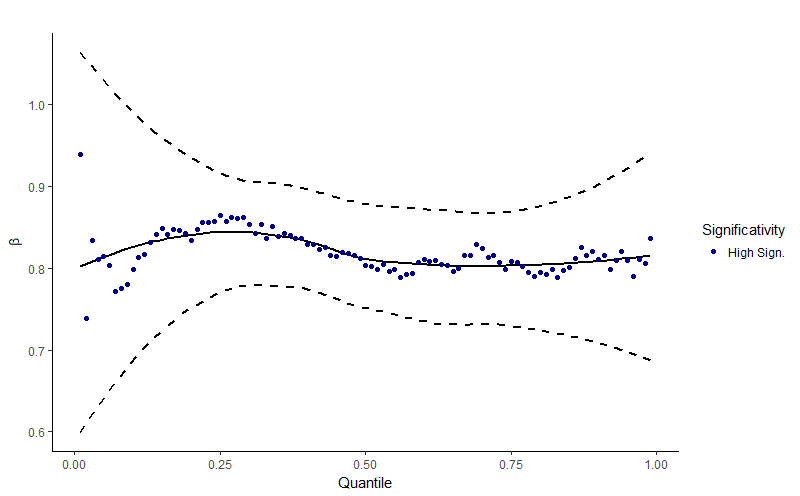}
		\caption{33 km}
	\end{subfigure}%
	\begin{subfigure}{.33\textwidth}
		\centering
		\includegraphics[width=.95\linewidth,height=5cm]{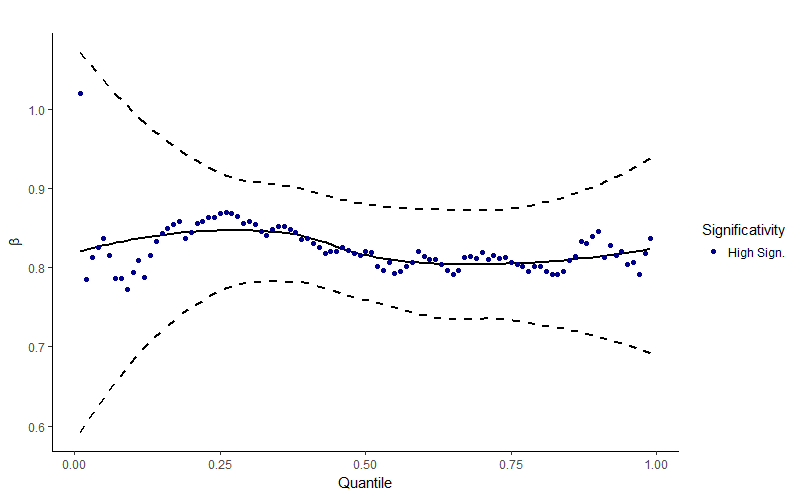}
		\caption{41 km}
	\end{subfigure}%
	\begin{subfigure}{.33\textwidth}
		\centering
		\includegraphics[width=.95\linewidth,height=5cm]{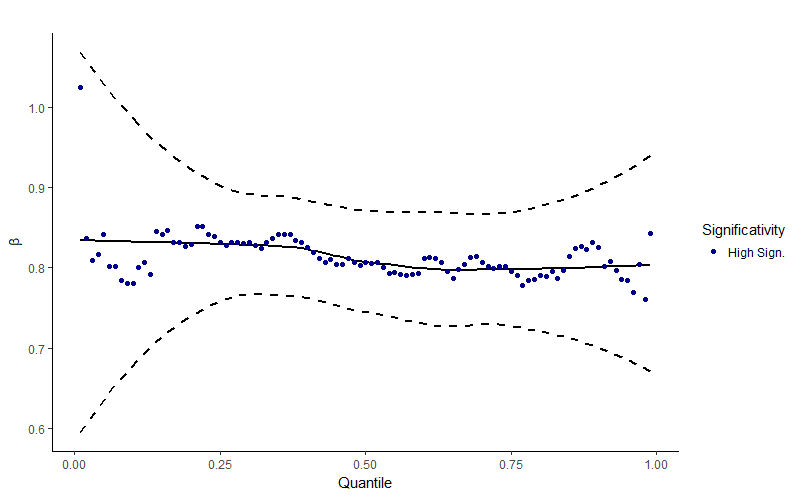}
		\caption{50 km}
\end{subfigure}

\caption*{\footnotesize {\textbf{Note:} Figure \ref{fig:sqr_l} shows the estimates of the Spatial AR model for every quantile. Solid line represents the smoothed function of the estimates, while dashed lines are the confidence interval at 95\%. Statistical significance is reported by different colours: Dark Blue= 0.01, Light Blue= 0.05, Red no significance.}}
\end{figure}

\begin{figure}[ht]
	\centering
	\caption{Spatial Quantile Regression for Land}
	\label{fig:sqr_g}
\begin{subfigure}{.33\textwidth}
	\centering
	\includegraphics[width=.95\linewidth,height=5cm]{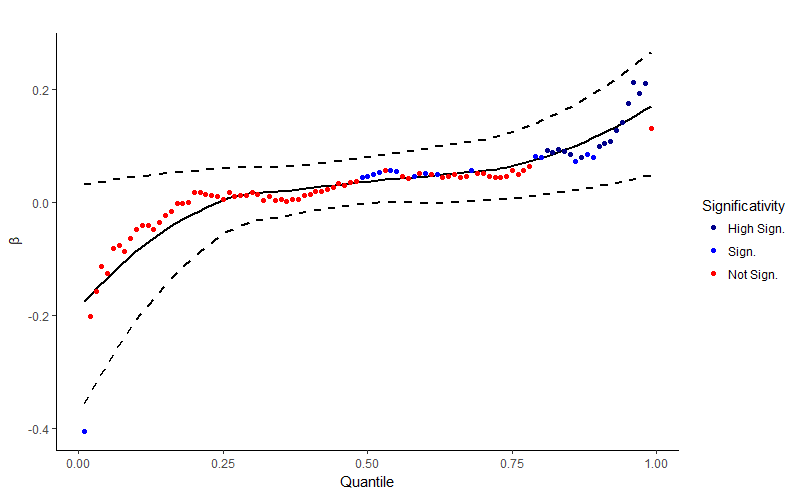}
	\caption{33 km}
\end{subfigure}%
\begin{subfigure}{.33\textwidth}
	\centering
	\includegraphics[width=.95\linewidth,height=5cm]{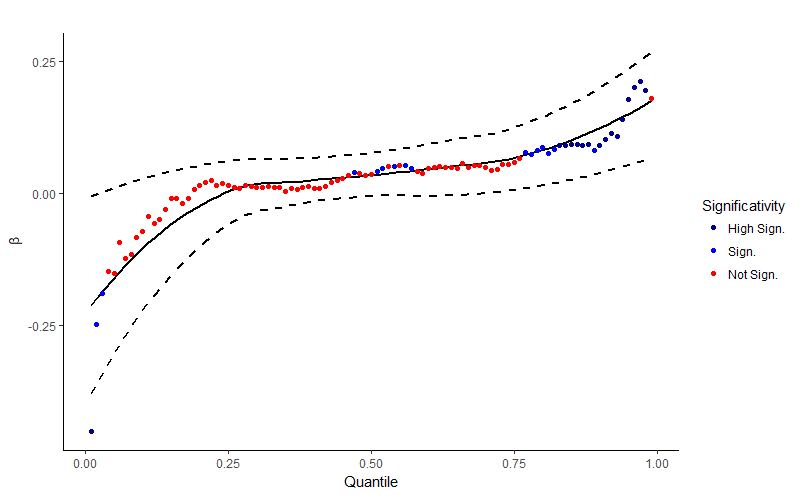}
	\caption{41 km}
\end{subfigure}%
\begin{subfigure}{.33\textwidth}
	\centering
	\includegraphics[width=.95\linewidth,height=5cm]{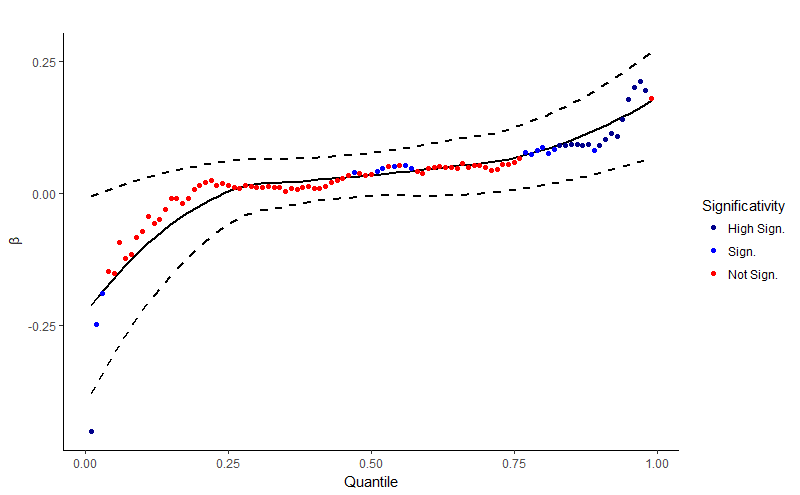}
	\caption{50 km}
\end{subfigure}

\caption*{\footnotesize {\textbf{Note:} Figure \ref{fig:sqr_g} shows the estimates of the Spatial AR model for every quantile. Solid line represents the smoothed function of the estimates, while dashed lines are the confidence interval at 95\%. Statistical significance is reported by different colours: Dark Blue= 0.01, Light Blue= 0.05, Red no significance.}}
\end{figure}

\begin{figure}[ht]
	\centering
	\caption{Spatial Quantile Regression for Subsides}
	\label{fig:sqr_s}
\begin{subfigure}{.33\textwidth}
	\centering
	\includegraphics[width=.95\linewidth,height=5cm]{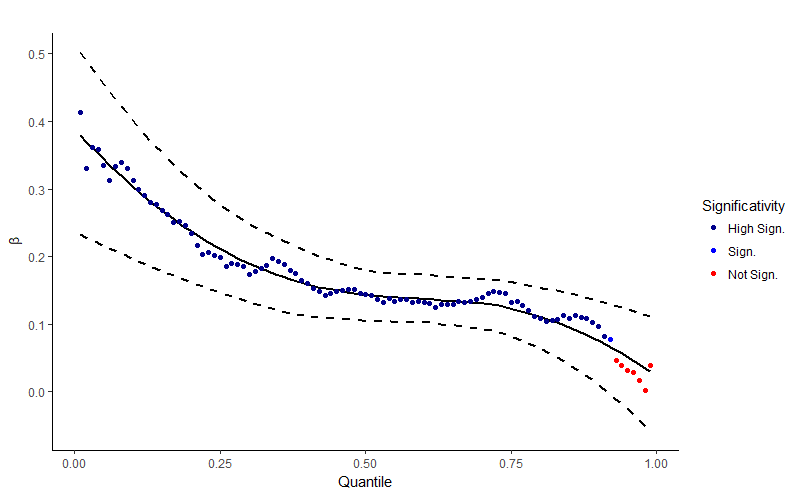}
	\caption{33 km}
\end{subfigure}%
\begin{subfigure}{.33\textwidth}
	\centering
	\includegraphics[width=.95\linewidth,height=5cm]{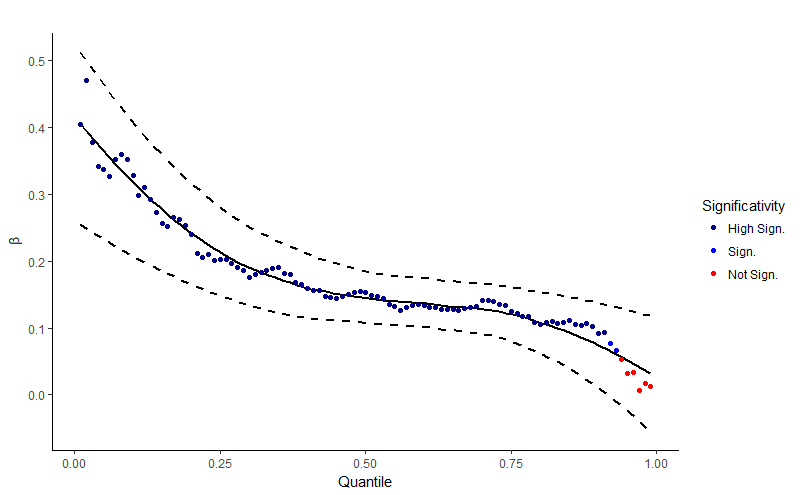}
	\caption{41 km}
\end{subfigure}%
\begin{subfigure}{.33\textwidth}
	\centering
	\includegraphics[width=.95\linewidth,height=5cm]{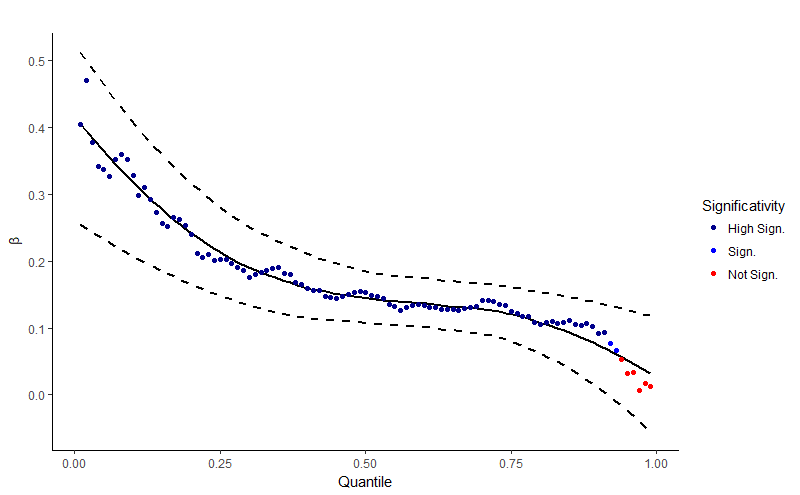}
	\caption{50 km}
\end{subfigure}

\caption*{\footnotesize {\textbf{Note:} Figure \ref{fig:sqr_s} shows the estimates of the Spatial AR model for every quantile. Solid line represents the smoothed function of the estimates, while dashed lines are the confidence interval at 95\%. Statistical significance is reported by different colours: Dark Blue= 0.01, Light Blue= 0.05, Red no significance.}}
\end{figure}

\begin{figure}[ht]
	\centering
	\caption{Spatial Quantile Regression for WY}
	\label{fig:sqr_wy}
	\begin{subfigure}{.33\textwidth}
		\centering
		\includegraphics[width=.95\linewidth,height=5cm]{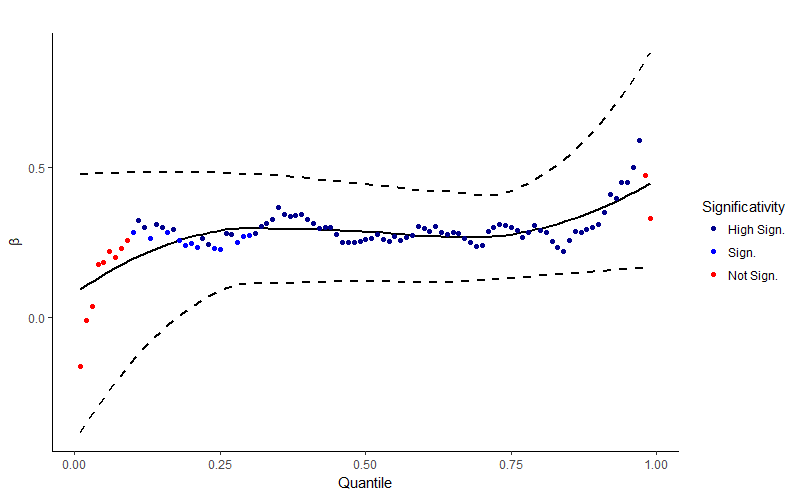}
		\caption{33 km}
	\end{subfigure}%
	\begin{subfigure}{.33\textwidth}
		\centering
		\includegraphics[width=.95\linewidth,height=5cm]{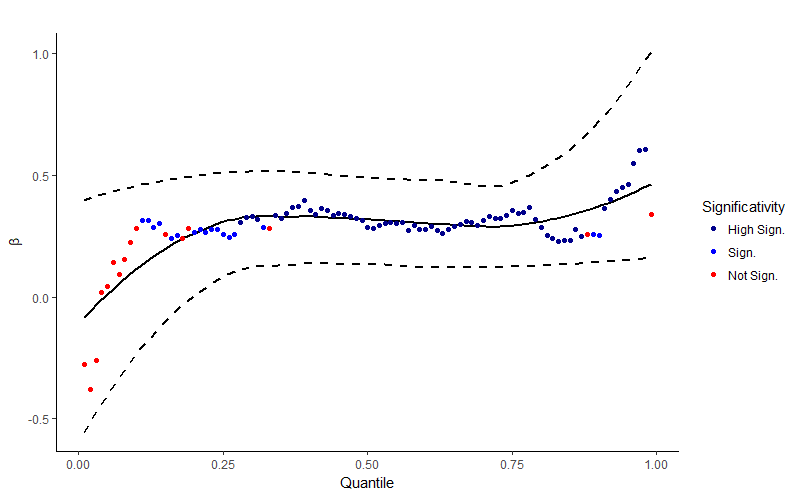}
		\caption{41 km}
	\end{subfigure}%
	\begin{subfigure}{.33\textwidth}
		\centering
		\includegraphics[width=.95\linewidth,height=5cm]{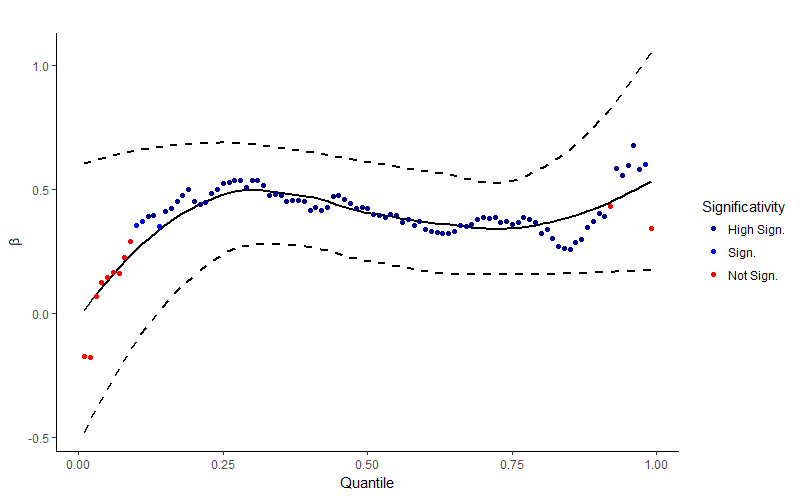}
		\caption{50 km}
	\end{subfigure}

\caption*{\footnotesize {\textbf{Note:} Figure \ref{fig:sqr_wy} shows the estimates of the Spatial AR model for every quantile. Solid line represents the smoothed function of the estimates, while dashed lines are the confidence interval at 95\%. Statistical significance is reported by different colours: Dark Blue= 0.01, Light Blue= 0.05, Red no significance.}}
\end{figure}

Quantile regression is an important method for including heterogeneous effects of covariates on a response variable \citep{koenker2001hallock}. However, as previously discussed economic performances of Italian farms are results of individual spatial interactions, making standard QR inference invalid. To include the presence of interactions, the quantile regression generalisation of the (linear) spatial lag model could be written as:
\begin{equation}
\label{eq:sqr}
Y =\rho(\tau) WY+ XB(\tau) + u
\end{equation}
where $Y=Q_{(\tau)}(Y|X)$ is the conditional quantile function of Y, $\tau$ refers to the selected quantile and $B(\tau)$ is the vector of the sensitivity coefficients of the conditional quantile on changes in value of the covariates  $X$. Estimating spatial quantile regression for different quantiles allows to predict the distribution of the outcome variable at given values of the explanatory variables \citep{mcmillen2017decompositions}. Furthermore, equation (\ref{eq:sqr}) argues that the spatial parameter, $\rho$, is dependent from the considered quantile $\tau$, allowing for different degree of spatial dependence across the conditional distribution. Further advantages of using a quantile regression approach are characteristically robust to the presence of outlier and heavy tailed distributions \citep{buchinsky1994qr,yu2003quantile,koenker2005book}.\\
The estimation procedures for quantile regression models can be classified in two distinct approaches. \cite{kim2004two} develop a two-stage quantile regression estimators\footnote{Literature refers to this method as ''fitted value'' approach and applications in a spatial framework can be found in \cite{zietz2008determinants},\cite{liao2012hedonic} and \cite{kostov2013choosing}.}, while \cite{chernozhukov2006instrumental} propose a generalisation of the instrumental variables framework to allow for estimation of quantile models\footnote{An in-depth analysis on this procedure is in \cite{yang2007su}, \cite{kostov2009spatial} and \cite{trzpiot2012spatial}.}. Both the approaches were initially developed to control for endogeneity in ''traditional'' quantile regression model, but can be easily adapted to deal with the spatial endogeneity in a Spatial Autoregressive quantile model.
The estimation procedure followed in this paper is based on the two-stage procedure in \cite{kim2004two} \footnote{To provide robustness to our results, in the next section we estimate our model following the approach developed by \cite{chernozhukov2006instrumental}.}.\\
On the first step, a variable constituting by the spatial lag of Y (in our case Added Value) is regressed over a set of instruments, as in Equation (\ref{eq:f_sqr}):
\begin{align}
\begin{split}
\label{eq:f_sqr}
\widehat{WY} &= Z\theta (\tau)+u\\
Z &=[X, WX]
\end{split}
\end{align}
The choice of the instruments follows the intuition in \cite{kelejan1998}, which demonstrate the consistency of this set of instruments. At the second stage, the variable $\widehat{WY}$ is added on a quantile regression of Y on the X's. In this way, we estimate an equation of the form:
\begin{equation}
\label{eq:s_sqr}
Y =\rho (\tau) \widehat{WY}+ XB (\tau)+ u
\end{equation}
Clearly, $\tau$ represents the same quantile in both equations (\ref{eq:f_sqr}) and (\ref{eq:s_sqr}). The consistency in this approach is guaranteed by estimating differentiated first stages for every quantile considered. Inference based solely on the second-stage of the procedure can be invalid. For this reason, standard errors for the overall two-stage procedure are bootstraped\footnote{New samples are constructed by drawing with replacement from the rows of the data frame holding y, WY, X, and Z. Both stages are re-estimated n-times using the series of bootstrap samples.}.
\begin{center}
	INSERT TABLE \ref{tab:sqr} AND FIGURES \ref{fig:sqr_k}, \ref{fig:sqr_l}, \ref{fig:sqr_g},\ref{fig:sqr_s},\ref{fig:sqr_wy}
\end{center}
Results of the spatial autoregressive quantile regression are presented in Table \ref{tab:sqr} and Figure from \ref{fig:sqr_k} to \ref{fig:sqr_wy}. While in Table \ref{tab:sqr} we report only the estimates for the extremes (0.1 and 0.9) and the quartiles (0.25, 0.50 and 0.75) of the distribution, Figures represent the entire conditional distribution of the parameters (i.e. every percentile between 0.01 and 0.99). The coefficient estimates for all variables are plotted together with their 95\% confidence bounds, while we omit the estimates for the intercept since it is not easily interpretable.\\
In overall, results are in line with previous estimates, even if decomposing the distribution of the effects offers some interesting insights. Labour is still the major components in fostering economic performances with a range between 0.85 and 0.8 and presents an higher level of significance (p-value< 0.01) . The distribution across the quantiles is pretty stationary, highlighting the independence of labour from the level of economic performances. Interestingly, fixed capital matters more in the extremes of the distribution by presenting a decreasing shape with a maximum in lowest quantile, which turn to increase at the first quartile (i.e. low and high  levels of fixed capital influence more economic performances).\\
However, the most interesting considerations are linked to public capital. This component shows a decreasing shape across all the distribution, with an inflection point in the neighbourhood of the median. Surprisingly, lower levels of subsidies have a greater impact on farm's performances (+1\% of public funding contributes to an increase of 0.4 \% in added value), while for the upper tail decrease to less than 0.1 and switch to be not significant. Land follows an increasing distributional shape, exhibiting a negative and not meaningful parameter until the median, while for higher quantiles the estimates become positive and significant. In other words, an increase of agricultural land has an impact on performances only for the farms with a wide initial UAA\footnote{UAA is an acronym for Utilised Agricultural Land.}.\\
Lastly, evidences of significant spillover effects are found. Distributional shape of $\rho$ parameter shows a positive and significant effect on economic performances, even if both lower and upper tails are not meaningful. This results provides clear evidences in favour of the existence of spatial patterns on agricultural activities in Italian lagged regions\footnote{For sake of clarity, we can think to farms A and B which are in the same neighbourhood. An increase in economic performances of farm A (resp.B) fosters value added also in B (resp. A). By consequence, sharing the benefits from neighbouring farms can determine the occurrence of spatial patterns.}.\\
As previously explained, to identify and evaluate direct and indirect effects a further step is necessary. Indeed, \cite{lesage2009introduction} shows how marginal impact of neighbouring areas become a composite parameter of the spatial weight matrix, $\rho$ and the estimated parameter\footnote{See Eq. \ref{eq:sar} in Section 4.2 .}. Spatial quantile regression requires an in-depth analysis for every quantiles considered. Table \ref{tab:marg_sqr} resume the decomposition of the marginal effects for the tails and the quartiles of the conditional distribution.

\begin{center}
	INSERT TABLE \ref{tab:marg_sqr}
\end{center}

Direct and total effects estimates are positive and significant across all the conditional distribution for all the variables, while results on land are ambigous and negligible. Nonetheless, labour and fixed capital provides homogenous parameter in changing the outcome variable conditional to different level of the covariates, we provide evidences of heterogeneous effects for the subsidies. In detail, the effect of the policies slightly decline for higher quantiles. This point is of particular interest in our analysis and will be discussed in conclusive section. \\
Looking at the indirect effects, capital becomes less effective and seems to be not linked to neighbouring characteristics, while positive and significant spillover effects are found in terms of human and public capital. However, these effects are limited to the inner part of the distribution. \\
\section{Robustness Check}
To resume, the approach presented in previous section is based on a spatial weight matrix with a cut-off distance equal to 33 km, which is the minimum distance for whom every units has at least one neighbour (i.e. absence of island). In other words, we impose a restriction on the possible spatial extension of the indirect effects. While, it is difficult to track the real extent of these effects, imposing a limited cut-off distance it seems to be reasonable. Indeed, Italian agricultural farms are characterized by a limited average dimension. The considered inputs are ''place-based'' and deeply embedded to local areas. In way of example, we can think to labour component which demonstrates a lower propensity to move, in comparison with secondary and tertiary sectors. In this sense, a cut-off equal to 33 km allows to limit our analysis to inter-municipal spillovers.\\
However, a traditional robustness check in spatial econometrics consists of considering different spatial weight matrices specification in estimating the models. For this reason, we propose an in-depth analysis for two different cut-off distances: 41 and 50 km \footnote{See figures \ref{fig:sqr_k}, \ref{fig:sqr_l}, \ref{fig:sqr_g},\ref{fig:sqr_s},\ref{fig:sqr_wy}}. 
While the estimates for a cut-off equal to 41 km are qualitatively and quantitatively in line with the baseline approach presented in previous sections and provide robustness to our results, some differences are found in the intensity of indirect effects for higher distances. Indeed, for a 50 km cut-off the impact of neighbouring performances becomes wider at the 1st quantile and decrease between 1st and 3rd quantiles. In other words, an increase of the cut-off distance enlarges also the number of farms in own neighbourhood and allows to consider an additional impact due to different agricultural and normative environments\footnote{Keeping a cut-off distance equal to 50 km can be considered as a strong robustness check. Stressing the cut-off distance to 50 km confirms the distributional shape of  $\rho$ along the conditional distribution, while the higher extent of the parameter does not invalidate the results of our analysis.}. Consequently, the greatest competitiveness on local market produces higher spillover only if neighbouring farms have low or very high performances. In other terms, we can think to the occurrence of low-low or high-high clusters.\\
Considering the decomposition in marginal effects, results show that the higher $\rho$ parameter is reflected in larger indirect effects. This results is in line with our expectation for a twofold reason. In first instance, we amplify cut-off distance to consider inter-provincial feedback on agricultural activities. Furthermore, the lack of significance highlight a narrow spatial frame (i.e. inter-municipal or intra-provincial) in which, potentially, the benefits can be shared.\\
An additional robustness check is the estimation of the Spatial Quantile Model following the approach in \cite{chernozhukov2006instrumental}. Their approach, based on a GMM procedure, use the predicted values of $\widehat{WY}$ obtained from an OLS regression of WY over a set of instruments (the same used in previous section). This instrumental variable is included as a covariate for a quantile regressions of Y - $\rho$ WY on X and $\widehat{WY}$. The estimated value of $\widehat{\rho}$ is the value that leads the coefficient on $\widehat{WY}$ to be closest to zero. After finding $\widehat{\rho}$, the values of $\beta$ are obtained by a quantile regression of $Y -\widehat{\rho} WY$ on X.\\
While standard errors in \cite{kim2004two} are calculated by bootstrap and it is computationally simpler, \cite{chernozhukov2006instrumental} provides a a direct formulation of standard errors and ensures robust finite sample performance \citep{kostov2009spatial}. In this sense, the wide sample size of our dataset makes preferable the approach in \cite{kim2004two}, while a robustness check following \cite{chernozhukov2006instrumental} procedure is needed.

\begin{center}
	INSERT TABLE \ref{tab:rob_sqr_CH} 
\end{center}

The estimates obtained with \cite{chernozhukov2006instrumental} approach are in line with the one presented in previous section. Minor differences are found in terms of the extent of $\rho$ especially for higher distances. However, in overall results provide evidences in favour of the correctness of our analysis.

\section{Conclusion}
This paper, analysing the case of the farms located in some southern regions of Italy, provides evidences in favour of the role played by different forms of capital (Human, Fixed, Land and Public) in fostering economic performances. Using an augmented production function we develop a multi-stage analysis. In the first step, we look at baseline results by estimating a linear model. In the second stage, we test for the presence of spatial autocorrelation or heterogeneity to estimate the occurrence of global or local spillovers. After providing evidences in favour of significant spatial spillovers, we implement a Spatial Autoregressive Quantile Regression Approach. Opening to a spatial regression approach allows to check for the presence of outliers and provides robust estimates. Furthermore, this approach makes possible to analyse the marginal impact of different forms of capital in fostering economic performances.\\
Spatial quantile estimates show clear evidences in favour of the occurrence of spatial spillovers. The positive and significant $\rho$ parameter across the quantile distribution highlights the presence of a strong spatial polarization of agricultural activities and performances. Clearly, these results can be influenced not only by the efficiency of the farms, but a key feature can be the sharing of similar environmental and weather characteristics. Furthermore, this paper provides opposite results to the ones of \cite{deca2017rica}. Aforementioned authors, considering farms located in Italy for the period between 2008 and 2009, demonstrate negative indirect effects. However, differences between the two paper are twofold. On one hand, \cite{deca2017rica} considers a time frame highly conditioned by macro-economic crisis, while omitting 2009 from our analysis allows to exclude the deepening year of the crisis. On the other hand, in this work we look only at farms located in Italian lagged regions. The limited spatial extension of this work provides a more balanced territorial framework and, in particular, allows to consider only the farms located in Italian less developed regions.\\
This point is the central pillar of our paper. Indeed, we demonstrate how public subsidies have a positive and significant marginal impact on economic performances. However, the intensity of the effects is 4 time lower than labour component. This assumption has a clear policy implication. Actually, Common Agricultural Policy (CAP) is designed as an instrument not linked to the production, but it aims to sustain farm's income and stabilise market prices.\\
A natural extension to improve the effectiveness of actual CAP can be the development of labour-market oriented policy. In this way, fostering labour participation to agricultural activities can substantially improve farm's performances. This approach can provide an additional channels to the lack of employment, especially in lagged regions, and contributes to increase the relative wealth of agriculture to overall national economy. Clearly, improving the efficiency of agricultural policies can also be a potential instrument in reducing economic gap between peripheral and core regions. However, this process is not straightforward and structural reforms are needed to promote the development of correct niche of specialization and, in overall, sustainability of agricultural sector.     
   
\bibliographystyle{apalike}
\bibliography{Di_Gennaro_De_Castris}

\begin{thebibliography}{}

\bibitem[Alston and James, 2002]{alston2002incidence}
Alston, J. and James, J.~S. (2002).
\newblock The incidence of agricultural policy.
\newblock In Gardner, B.~L. and Rausser, G.~C., editors, {\em Handbook of
  Agricultural Economics}, volume 2, Part 2, chapter~33, pages 1689--1749.
  Elsevier, 1 edition.

\bibitem[Anselin, 1995]{anselin1995local}
Anselin, L. (1995).
\newblock {L}ocal {I}ndicators of {S}patial {A}ssociation- {LISA}.
\newblock {\em Geographical Analysis}, 27(2):93--115.

\bibitem[Anselin, 2003]{anselin2003spatial}
Anselin, L. (2003).
\newblock {S}patial {E}xternalities, {S}patial {M}ultipliers and {S}patial
  {E}conometrics.
\newblock {\em International Regional Science Review}, 26(2):153--166.

\bibitem[Anselin et~al., 1996]{anselin1996simple}
Anselin, L., Bera, A.~K., Florax, R., and Yoon, M.~J. (1996).
\newblock Simple diagnostic tests for spatial dependence.
\newblock {\em Regional Science and Urban Economics}, 26(1):77 -- 104.

\bibitem[Anselin and Le~Gallo, 2006]{anselin2006interpolation}
Anselin, L. and Le~Gallo, J. (2006).
\newblock {I}nterpolation of {A}ir {Q}uality {M}easures in {H}edonic {H}ouse
  {P}rice {M}odels: {S}patial {A}spects.
\newblock {\em Spatial Economic Analysis}, 1(1):31--52.

\bibitem[Arraiz et~al., 2010]{arraiz2010spatial}
Arraiz, I., Drukker, D.~M., Kelejian, H.~H., and Prucha, I.~R. (2010).
\newblock A {S}patial {C}liff-{O}rd-{t}ype model with heteroskedastic
  innovations: small and large sample results.
\newblock {\em Journal of Regional Science}, 50(2):592--614.

\bibitem[Autant-Bernard, 2012]{autant2012spatial}
Autant-Bernard, C. (2012).
\newblock {S}patial {E}conometrics of {I}nnovation: {R}ecent {C}ontributions
  and {R}esearch {P}erspectives.
\newblock {\em Spatial Economic Analysis}, 7(4):403--419.

\bibitem[Bill\'{e} et~al., 2015]{bille2015spatial}
Bill\'{e}, A.~G., Salvioni, C., and Benedetti, R. (2015).
\newblock {S}patial {H}eterogeneity in {P}roduction {F}unctions {M}odels.
\newblock 150th seminar, {O}ctober 22-23, 2015, {E}dinburgh, {S}cotland,
  European Association of Agricultural Economists.

\bibitem[Bojnec and Latruffe, 2013]{bojnec2013farm}
Bojnec, {\v{S}}. and Latruffe, L. (2013).
\newblock Farm size, agricultural subsidies and farm performance in {S}lovenia.
\newblock {\em Land Use Policy}, 32:207 -- 217.

\bibitem[Buchinsky, 1994]{buchinsky1994qr}
Buchinsky, M. (1994).
\newblock {C}hanges in the {U.S. W}age {S}tructure 1963-1987: {A}pplication of
  {Q}uantile {R}egression.
\newblock {\em Econometrica}, 62(2):405--458.

\bibitem[Chernozhukov and Hansen, 2006]{chernozhukov2006instrumental}
Chernozhukov, V. and Hansen, C. (2006).
\newblock Instrumental quantile regression inference for structural and
  treatment effect models.
\newblock {\em Journal of Econometrics}, 132(2):491 -- 525.

\bibitem[Daugbjerg, 2003]{daugbjerg2003policy}
Daugbjerg, C. (2003).
\newblock {Policy feedback and paradigm shift in EU agricultural policy: the
  effects of the MacSharry reform on future reform}.
\newblock {\em Journal of European Public Policy}, 10(3):421--437.

\bibitem[De~Castris and Di~Gennaro, 2017]{deca2017rica}
De~Castris, M. and Di~Gennaro, D. (2017).
\newblock {What is Below the CAP? Evaluating Spatial Patterns in Agricultural
  Subsidies}.
\newblock In {\em XXXVIII Annual Scientific Conference of the A. I. S. Re.
  Italian Association of Regional Science, Cagliari (CA), September 20-22,
  2017}.

\bibitem[De~Dominicis et~al., 2013]{de2013regional}
De~Dominicis, L., Florax, R.~J., and De~Groot, H.~L. (2013).
\newblock {Regional clusters of innovative activity in Europe: are social
  capital and geographical proximity key determinants}?
\newblock {\em Applied Economics}, 45(17):2325--2335.

\bibitem[Dwyer et~al., 2007]{dwyer2007european}
Dwyer, J., Ward, N., Lowe, P., and Baldock, D. (2007).
\newblock European rural development under the common agricultural policy's
  ''second pillar'': Institutional conservatism and innovation.
\newblock {\em Regional Studies}, 41(7):873--888.

\bibitem[Fader et~al., 2013]{fader2013spatial}
Fader, M., Gerten, D., Krause, M., Lucht, W., and Cramer, W. (2013).
\newblock Spatial decoupling of agricultural production and consumption:
  quantifying dependences of countries on food imports due to domestic land and
  water constraints.
\newblock {\em Environmental Research Letters}, 8(1):014046.

\bibitem[Fleischer and Tchetchik, 2005]{fleischer2005does}
Fleischer, A. and Tchetchik, A. (2005).
\newblock Does rural tourism benefit from agriculture?
\newblock {\em Tourism Management}, 26(4):493 -- 501.

\bibitem[Folmer et~al., 1995]{folmer2013common}
Folmer, C., Keyzer, M.~A., Merbis, M.~D., Stolwijk, H.~J., and Veenendaal,
  P.~J. (1995).
\newblock {\em The common agricultural policy beyond the {M}ac{S}harry reform},
  volume 230 of {\em Contribution to Economic Analysis}.
\newblock Elsevier.

\bibitem[Giannakis and Bruggeman, 2015]{giannakis2015highly}
Giannakis, E. and Bruggeman, A. (2015).
\newblock The highly variable economic performance of {E}uropean agriculture.
\newblock {\em Land Use Policy}, 45(Supplement C):26 -- 35.

\bibitem[Gorton et~al., 2009]{gorton2009folly}
Gorton, M., Hubbard, C., and Hubbard, L. (2009).
\newblock {The Folly of European Union Policy Transfer: Why the Common
  Agricultural Policy (CAP) Does Not Fit Central and Eastern Europe}.
\newblock {\em Regional Studies}, 43(10):1305--1317.

\bibitem[Griliches, 1964]{griliches1964research}
Griliches, Z. (1964).
\newblock {Research Expenditures, Education, and the Aggregate Agricultural
  Production Function}.
\newblock {\em The American Economic Review}, 54(6):961--974.

\bibitem[Hennessy, 1998]{hennessy1998production}
Hennessy, D.~A. (1998).
\newblock {The Production Effects of Agricultural Income Support Policies under
  Uncertainty}.
\newblock {\em American Journal of Agricultural Economics}, 80(1):46--57.

\bibitem[Herrendorf et~al., 2015]{herrendorf2015sectoral}
Herrendorf, B., Herrington, C., and Valentinyi, A. (2015).
\newblock {Sectoral Technology and Structural Transformation}.
\newblock {\em American Economic Journal: Macroeconomics}, 7(4):104--33.

\bibitem[INEA, 2008]{inea2008}
INEA (2008).
\newblock {\em Annuario dell' agricoltura Italiana}, volume~63.
\newblock INEA.

\bibitem[Kazukauskas et~al., 2014]{kazukauskas2014impact}
Kazukauskas, A., Newman, C., and Sauer, J. (2014).
\newblock The impact of decoupled subsidies on productivity in agriculture: a
  cross-country analysis using microdata.
\newblock {\em Agricultural Economics}, 45(3):327--336.

\bibitem[Kazukauskas et~al., 2010]{kazukauskas2010analysing}
Kazukauskas, A., Newman, C.~F., and Thorne, F.~S. (2010).
\newblock {Analysing the Effect of Decoupling on Agricultural Production:
  Evidence from Irish Dairy Farms using the Olley and Pakes Approach}.
\newblock {\em Journal of International Agricultural Trade and Development},
  59(3).

\bibitem[Kelejian and Prucha, 1998]{kelejan1998}
Kelejian, H.~H. and Prucha, I. (1998).
\newblock {A Generalized Spatial Two-Stage Least Squares Procedure for
  Estimating a Spatial Autoregressive Model with Autoregressive Disturbances}.
\newblock {\em The Journal of Real Estate Finance and Economics},
  17(1):99--121.

\bibitem[Kelejian and Prucha, 2010]{kelejian2010specification}
Kelejian, H.~H. and Prucha, I.~R. (2010).
\newblock Specification and estimation of spatial autoregressive models with
  autoregressive and heteroskedastic disturbances.
\newblock {\em Journal of Econometrics}, 157(1):53 -- 67.
\newblock Nonlinear and Nonparametric Methods in Econometrics.

\bibitem[Kim et~al., 2003]{kim2003measuring}
Kim, C.~W., Phipps, T.~T., and Anselin, L. (2003).
\newblock Measuring the benefits of air quality improvement: a spatial hedonic
  approach.
\newblock {\em Journal of Environmental Economics and Management}, 45(1):24 --
  39.

\bibitem[Kim and Muller, 2004]{kim2004two}
Kim, T.-H. and Muller, C. (2004).
\newblock Two-stage quantile regression when the first stage is based on
  quantile regression.
\newblock {\em Econometrics Journal}, 7(1):218--231.

\bibitem[Koenker, 2005]{koenker2005book}
Koenker, R. (2005).
\newblock {\em {Quantile Regression}}.
\newblock Cambridge University Press.

\bibitem[Koenker and Hallock, 2001]{koenker2001hallock}
Koenker, R. and Hallock, K. (2001).
\newblock {Quantile Regression}.
\newblock {\em Journal of Economic Perspectives}, 15(4):143--156.

\bibitem[Kostov, 2009]{kostov2009spatial}
Kostov, P. (2009).
\newblock {A Spatial Quantile Regression Hedonic Model of Agricultural Land
  Prices}.
\newblock {\em Spatial Economic Analysis}, 4(1):53--72.

\bibitem[Kostov, 2013]{kostov2013choosing}
Kostov, P. (2013).
\newblock Choosing the right spatial weighting matrix in a quantile regression
  model.
\newblock {\em ISRN Economics}, 2013.

\bibitem[LeSage and Pace, 2009]{lesage2009introduction}
LeSage, J. and Pace, R.~K. (2009).
\newblock {\em Introduction to {S}patial {E}conometrics}.
\newblock CRC Press.

\bibitem[Liao and Wang, 2012]{liao2012hedonic}
Liao, W.-C. and Wang, X. (2012).
\newblock Hedonic house prices and spatial quantile regression.
\newblock {\em Journal of Housing Economics}, 21(1):16 -- 27.

\bibitem[Lucas, 2004]{lucas2004life}
Lucas, Jr, R.~E. (2004).
\newblock {Life Earnings and Rural-Urban Migration}.
\newblock {\em Journal of Political Economy}, 112(S1):S29--S59.

\bibitem[Marrocu et~al., 2013]{marrocu2011proximity}
Marrocu, E., Paci, R., and Usai, S. (2013).
\newblock Proximity, networking and knowledge production in {E}urope: {W}hat
  lessons for innovation policy?
\newblock {\em Technological Forecasting and Social Change}, 80(8):1484--1498.

\bibitem[Martin and Mitra, 2001]{martin2001productivity}
Martin, W. and Mitra, D. (2001).
\newblock {Productivity Growth and Convergence in Agriculture versus
  Manufacturing}.
\newblock {\em Economic Development and Cultural Change}, 49(2):403--422.

\bibitem[McMillen and Shimizu, 2017]{mcmillen2017decompositions}
McMillen, D. and Shimizu, C. (2017).
\newblock Decompositions of spatially varying quantile distribution estimates:
  The rise and fall of tokyo house prices.
\newblock {\em Urbana}, 51:61801.

\bibitem[Minviel and Latruffe, 2017]{minviel2017effect}
Minviel, J.~J. and Latruffe, L. (2017).
\newblock Effect of public subsidies on farm technical efficiency: a
  meta-analysis of empirical results.
\newblock {\em Applied Economics}, 49(2):213--226.

\bibitem[Olper et~al., 2014]{olper2014cap}
Olper, A., Raimondi, V., Cavicchioli, D., and Vigani, M. (2014).
\newblock {Do CAP payments reduce farm labour migration? A panel data analysis
  across EU regions}.
\newblock {\em European Review of Agricultural Economics}, 41(5):843--873.

\bibitem[Pe'er et~al., 2014]{pe2014eu}
Pe'er, G., Dicks, L., Visconti, P., Arlettaz, R., B{\'a}ldi, A., Benton, T.,
  Collins, S., Dieterich, M., Gregory, R., and Hartig, F. (2014).
\newblock {EU} agricultural reform fails on biodiversity.
\newblock {\em Science}, 344(6188):1090--1092.

\bibitem[Porkka et~al., 2013]{porkka2013food}
Porkka, M., Kummu, M., Siebert, S., and Varis, O. (2013).
\newblock {From Food Insufficiency towards Trade Dependency: A Historical
  Analysis of Global Food Availability}.
\newblock {\em PLoS ONE}, 8(12):e82714.

\bibitem[Rizov et~al., 2013]{rizov2013cap}
Rizov, M., Pokrivcak, J., and Ciaian, P. (2013).
\newblock {CAP Subsidies and Productivity of the EU Farms}.
\newblock {\em Journal of Agricultural Economics}, 64(3):537--557.

\bibitem[Rodr\`{\i}guez-Pose and Fratesi, 2004]{rodriguez2004between}
Rodr\`{\i}guez-Pose, A. and Fratesi, U. (2004).
\newblock Between development and social policies: The impact of european
  structural funds in objective 1 regions.
\newblock {\em Regional Studies}, 38(1):97--113.

\bibitem[Trzpiot, 2012]{trzpiot2012spatial}
Trzpiot, G. (2012).
\newblock Spatial quantile regression.
\newblock {\em Comparative Economic Research}, 15(4):265--279.

\bibitem[Wegener et~al., 2011]{wegener2011administering}
Wegener, S., Labar, K., Petrick, M., Marquardt, D., Theesfeld, I., and
  Buchenrieder, G. (2011).
\newblock {Administering the Common Agricultural Policy in Bulgaria and
  Romania: obstacles to accountability and administrative capacity}.
\newblock {\em International Review of Administrative Sciences},
  77(3):583--608.

\bibitem[Yang and Su, 2007]{yang2007su}
Yang, Z. and Su, L. (2007).
\newblock {Instrumental Variable Quantile Estimation of Spatial Autoregressive
  Models}.
\newblock Working Papers 05-2007, Singapore Management University, School of
  Economics.

\bibitem[Yu et~al., 2003]{yu2003quantile}
Yu, K., Lu, Z., and Stander, J. (2003).
\newblock Quantile regression: applications and current research areas.
\newblock {\em Journal of the Royal Statistical Society: Series D (The
  Statistician)}, 52(3):331--350.

\bibitem[Zhu and Demeter, 2012]{zhu2012technical}
Zhu, X. and Demeter, R.~M. (2012).
\newblock {Technical efficiency and productivity differentials of dairy farms
  in three EU countries: the role of CAP subsidies}.
\newblock {\em Agricultural Economics Review}, 13(1).

\bibitem[Zietz et~al., 2008]{zietz2008determinants}
Zietz, J., Zietz, E.~N., and Sirmans, G.~S. (2008).
\newblock {Determinants of House Prices: A Quantile Regression Approach}.
\newblock {\em The Journal of Real Estate Finance and Economics},
  37(4):317--333.

\end{thebibliography}

\cleardoublepage
\appendix
\begin{center}
	\textbf{List of Tables and Figures}
\end{center}
\bigskip
\bigskip
\bigskip
\begin{table}[htbp!]
	\centering
	\caption{List of Variables}
	\small{
		\resizebox{\linewidth}{!}{
			
			\begin{tabular}{|l|c|c|l|}
				\hline
				Variable & Label& \multicolumn{1}{l|}{Measure unit} & Description \\
				\hline
				Value-added & VA    & \geneuro     & Total Revenues-Current Expenses \\
				\multicolumn{1}{|l|}{\multirow{2}[0]{*}{Labour  }} & \multirow{2}[0]{*}{L} & \multirow{2}[0]{*}{Unit} & \multirow{1}[0]{*}{Full time worker. Every 2200 working hours in the farm} \\
				&       &       & \multirow{1}[0]{*}{represent a FTW}\\
				Capital stock & K     &   \geneuro   & Land Capital+ Agricultural Fixed Capital \\
				Land  & G     & Hectares & Utilised Agricultural Area (UAA) \\
				Subsidies  & S     & \geneuro     & Total Amount of subsidies for farm \\
				\hline
			\end{tabular}%
			\label{tab:var}%
		}
	}
\end{table}%

\begin{table}[htbp]
	\centering
	\caption{Micro-Founded OLS EStimates}
	\noindent\resizebox{\linewidth}{!}{
	\begin{tabular}{|c|rl|rl|rl|rl|}
		\cline{2-9}    \multicolumn{1}{c|}{} & \multicolumn{2}{c|}{[1]} & \multicolumn{2}{c|}{[2]} & \multicolumn{2}{c|}{[3]} & \multicolumn{2}{c|}{[4]} \\
		\hline
		\multirow{2}[1]{*}{Constant} & 6.771 & ***   & 7.32  & ***   & 6.652 & ***   & 6.777 & *** \\
		& \multicolumn{1}{r}{[0.194]} &       & \multicolumn{1}{r}{[0.202]} &       & \multicolumn{1}{r}{[0.185]} &       & \multicolumn{1}{r}{[0.207]} &  \\
		\multirow{2}[0]{*}{K} & 0.271 & ***   & 0.191 & ***   & 0.162 & ***   & 0.155 & *** \\
		& \multicolumn{1}{r}{[0.016]} &       & \multicolumn{1}{r}{[0.019]} &       & \multicolumn{1}{r}{[0.018]} &       & \multicolumn{1}{r}{[0.019]} &  \\
		\multirow{2}[0]{*}{L} & 0.918 & ***   & 0.866 & ***   & 0.875 & ***   & 0.868 & *** \\
		& \multicolumn{1}{r}{[0.027]} &       & \multicolumn{1}{r}{[0.027]} &       & \multicolumn{1}{r}{[0.026]} &       & \multicolumn{1}{r}{[0.027]} &  \\
		\multirow{2}[0]{*}{G} & \multicolumn{2}{c|}{\multirow{2}[0]{*}{}} & 0.146 & ***   & \multicolumn{2}{c|}{\multirow{2}[0]{*}{}} & 0.03  &  \\
		& \multicolumn{2}{c|}{} & \multicolumn{1}{r}{[0.019]} &       & \multicolumn{2}{c|}{} & \multicolumn{1}{r}{[0.023]} &  \\
		\multirow{2}[0]{*}{S} & \multicolumn{2}{c|}{\multirow{2}[0]{*}{}} & \multicolumn{2}{c|}{\multirow{2}[0]{*}{}} & 0.169 & ***   & 0.154 & *** \\
		& \multicolumn{2}{c|}{} & \multicolumn{2}{c|}{} & \multicolumn{1}{r}{[0.014]} &       & \multicolumn{1}{r}{[0.018]} &  \\
		\multirow{2}[0]{*}{N.observations} & \multicolumn{2}{c|}{\multirow{2}[0]{*}{1298}} & \multicolumn{2}{c|}{\multirow{2}[0]{*}{1298}} & \multicolumn{2}{c|}{\multirow{2}[0]{*}{1298}} & \multicolumn{2}{c|}{\multirow{2}[0]{*}{1298}} \\
		& \multicolumn{2}{c|}{} & \multicolumn{2}{c|}{} & \multicolumn{2}{c|}{} & \multicolumn{2}{c|}{} \\
		\multirow{2}[1]{*}{Adjusted $R^{2}$} & \multicolumn{2}{c|}{\multirow{2}[1]{*}{0.666}} & \multicolumn{2}{c|}{\multirow{2}[1]{*}{0.681}} & \multicolumn{2}{c|}{\multirow{2}[1]{*}{0.698}} & \multicolumn{2}{c|}{\multirow{2}[1]{*}{0.698}} \\
		& \multicolumn{2}{c|}{} & \multicolumn{2}{c|}{} & \multicolumn{2}{c|}{} & \multicolumn{2}{c|}{} \\
		\hline
		Distance & \multicolumn{8}{c|}{SARMA Lagrange Multiplier Test for Spatial Dependence} \\
		\hline
		33 km & 101.669 & ***   & 131.948 & ***   & 101.288 & ***   & 106.451 & *** \\
		41 km & 123.992 & ***   & 172.304 & ***   & 126.178 & ***   & 134.31 & *** \\
		50 km & 132.44 & ***   & 183.657 & ***   & 141.623 & ***   & 149.228 & *** \\
		\hline
	\end{tabular}%
}
	\label{tab:ols}
\caption*{\footnotesize {\textbf{Note:} Table \ref{tab:ols} presents the results of OLS model for 2008. Estimates are considered in terms of elasticities, while standard errors are in square brackets. To test for spatial dependence we use a Lagrange Multiplier Test \citep{anselin1996simple}. The ambiguous results on Spatial Autoregressive and Spatial Error models, suggests to use a SARMA. Statistics of this test (for SARMA model) are reported by considering three distinct cutt-off: 33 km corresponding to the minimum distance for which every unit has at least one neighbour, 41 km (1.25* min. dist) and 50 km (1.5* min dist.).\\
Statistical significance: *** <0.001, ** 0.01, * 0.05, $^{\circ}$ 0.1 }}	
\end{table}%

\begin{table}[htbp]
	\centering
	\caption{SARMA Cobb-Douglas Estimation}
	\noindent\resizebox{\linewidth}{!}{
	\begin{tabular}{|c|llll|llll|llll|}
		\cline{2-13}    \multicolumn{1}{r|}{} & \multicolumn{4}{c|}{Distance=33 Km} & \multicolumn{4}{c|}{Distance=41 Km} & \multicolumn{4}{c|}{Distance=50 km} \\
		\cline{2-13}    \multicolumn{1}{r|}{} & \multicolumn{1}{c}{[1]} & \multicolumn{1}{c}{[2]} & \multicolumn{1}{c}{[3]} & \multicolumn{1}{c|}{[4]} & \multicolumn{1}{c}{[1]} & \multicolumn{1}{c}{[2]} & \multicolumn{1}{c}{[3]} & \multicolumn{1}{c|}{[4]} & \multicolumn{1}{c}{[1]} & \multicolumn{1}{c}{[2]} & \multicolumn{1}{c}{[3]} & \multicolumn{1}{c|}{[4]} \\
		\hline
		\multirow{2}[0]{*}{Constant} & 3.96*** & 4.16*** & 3.44*** & 3.55*** & 4.24*** & 4.5*** & 3.3*** & 3.55*** & 3.52** & 3.85** & 2.54* & 2.46* \\
		& [0.91] & [0.93] & [0.81] & [0.82] & [1.11] & [1.14] & [0.98] & [0.82] & [1.23] & [1.24] & [1.09] & [1.10] \\
		\multirow{2}[0]{*}{K} & 0.28*** & 0.2*** & 0.18*** & 0.17*** & 0.29*** & 0.2*** & 0.18*** & 0.17*** & 0.29*** & 0.21*** & 0.19*** & 0.18*** \\
		& [0.02] & [0.02] & [0.02] & [0.02] & [0.02] & [0.02] & [0.02] & [0.02] & [0.02] & [0.02] & [0.02] & [0.02] \\
		\multirow{2}[0]{*}{L} & 0.89*** & 0.82*** & 0.84*** & 0.82*** & 0.89*** & 0.82*** & 0.84*** & 0.82*** & 0.88*** & 0.81*** & 0.83*** & 0.82*** \\
		& [0.03] & [0.03] & [0.03] & [0.03] & [0.03] & [0.03] & [0.03] & [0.03] & [0.03] & [0.03] & [0.03] & [0.03] \\
		\multirow{2}[0]{*}{G} &       & 0.17*** &       & 0$0.05^{\circ}$  &       & 0.17*** &       & $0.05^{\circ}$  &       & 0.17*** &       & 0.05* \\
		&       & [0.02] &       & [0.02] &       & [0.02] &       & [0.02] &       & [0.02] &       & [0.02] \\
		\multirow{2}[0]{*}{S} &       &       & 0.17*** & 0.15*** &       &       & 0.17*** & 0.15*** &       &       & 0.17*** & 0.15*** \\
		&       &       & [0.02] & [0.02] &       &       & [0.02] & [0.02] &       &       & [0.02] & [0.02] \\
		\multirow{2}[0]{*}{$\rho$} & 0.26** & 0.29*** & 0.29*** & 0.3*** & 0.22* & 0.25* & 0.3** & 0.3*** & 0.29* & 0.31** & 0.37*** & 0.39*** \\
		& [0.08] & [0.08] & [0.07] & [0.07] & [0.10] & [0.10] & [0.09] & [0.07] & [0.11] & [0.11] & [0.10] & [0.10] \\
		\multirow{2}[1]{*}{$\lambda$} & 0.29** & 0.36*** & 0.25** & 0.26** & 0.42*** & 0.51*** & 0.36*** & 0.26** & 0.45*** & 0.54*** & 0.4*** & 0.41*** \\
		& [0.10] & [0.08] & [0.09] & [0.09] & [0.11] & [0.09] & [0.10] & [0.09] & [0.11] & [0.10] & [0.11] & [0.10] \\
		\hline
	\end{tabular}%
}
	\label{tab:sarma}
	\caption*{\footnotesize {\textbf{Note:} Table \ref{tab:sarma} presents the results of SARMA model for 2008. Estimates are considered in terms of elasticities, while standard errors are in square brackets. Three distinct cutt-off are considered: 33 km corresponding to the minimum distance for which every unit has at least one neighbour, 41 km (1.25* min. dist) and 50 km (1.5* min dist.).\\
			Statistical significance: *** <0.001, ** 0.01, * 0.05, $^{\circ}$ 0.1 }}
\end{table}%

\begin{table}[htbp]
	\centering
	\caption{Marginal Impact}
	\noindent\resizebox{\linewidth}{!}{
	\begin{tabular}{|c|c|rrr|rrr|lll|lll|}
		\cline{3-14}    \multicolumn{1}{r}{} &       & \multicolumn{3}{c|}{[1]} & \multicolumn{3}{c|}{[2]} & \multicolumn{3}{c|}{[3]} & \multicolumn{3}{c|}{[4]} \\
		\cline{3-14}    \multicolumn{1}{r}{} &       & \multicolumn{1}{c}{D} & \multicolumn{1}{c}{I} & \multicolumn{1}{c|}{T} & \multicolumn{1}{c}{D} & \multicolumn{1}{c}{I} & \multicolumn{1}{c|}{T} &  \multicolumn{1}{c}{D} & \multicolumn{1}{c}{I} & \multicolumn{1}{c|}{T} & \multicolumn{1}{c}{D} & \multicolumn{1}{c}{I} & \multicolumn{1}{c|}{T}\\
		\hline
		\multirow{8}[2]{*}{Cut-Off = 33 km} & \multirow{2}[1]{*}{K} & \multicolumn{1}{l}{0.29***} & \multicolumn{1}{l}{0.10*} & \multicolumn{1}{l|}{0.38***} & \multicolumn{1}{l}{0.20***} & \multicolumn{1}{l}{0.08*} & \multicolumn{1}{l|}{0.28***} & 0.18*** & 0.07** & 0.25*** & 0.17*** & 0.07* & 0.24*** \\
		&       & \multicolumn{1}{l}{[14.33]} & \multicolumn{1}{l}{[2.18]} & \multicolumn{1}{l|}{[6.83]} & \multicolumn{1}{l}{[9.44]} & \multicolumn{1}{l}{[2.27]} & \multicolumn{1}{l|}{[5.72]} & [8.74] & [2.65] & [6.17] & [7.98] & [2.53] & [5.67] \\
		& \multirow{2}[0]{*}{L} & \multicolumn{1}{l}{0.89***} & \multicolumn{1}{l}{0.30*} & \multicolumn{1}{l|}{1.19***} & \multicolumn{1}{l}{0.82***} & \multicolumn{1}{l}{0.33*} & \multicolumn{1}{l|}{1.15***} & 0.84*** & 0.34** & 1.18*** & 0.83*** & 0.35** & 1.18*** \\
		&       & \multicolumn{1}{l}{[29.99]} & \multicolumn{1}{l}{[2.28]} & \multicolumn{1}{l|}{[8.49]} & \multicolumn{1}{l}{[26.26]} & \multicolumn{1}{l}{[2.48]} & \multicolumn{1}{l|}{[8.11]} & [27.83] & [2.88] & [9.35] & [26.9] & [2.83] & [9.08] \\
		& \multirow{2}[0]{*}{G} & \multicolumn{1}{l}{} & \multicolumn{1}{l}{} & \multicolumn{1}{l|}{} & \multicolumn{1}{l}{0.17***} & \multicolumn{1}{l}{0.07*} & \multicolumn{1}{l|}{0.24***} &       &       &       & 0.05* & 0.02  & 0.07* \\
		&       &       &       &       & \multicolumn{1}{l}{[8.12]} & \multicolumn{1}{l}{[2.37]} & \multicolumn{1}{l|}{[5.87]} &       &       &       & [2.08] & [1.60] & [2.03] \\
		& \multirow{2}[1]{*}{S} & \multicolumn{1}{l}{} & \multicolumn{1}{l}{} & \multicolumn{1}{l|}{} & \multicolumn{1}{l}{} & \multicolumn{1}{l}{} & \multicolumn{1}{l|}{} & 0.17*** & 0.07** & 0.24*** & 0.15*** & 0.06** & 0.21*** \\
		&       &       &       &       &       &       &       & [10.77] & [2.79] & [7.26] & [7.61] & [2.67] & [5.91] \\
		\hline
		\multirow{8}[2]{*}{Cut-Off = 41 km} & \multirow{2}[1]{*}{K} & \multicolumn{1}{l}{0.29***} & \multicolumn{1}{l}{$0.08^{\circ}$} & \multicolumn{1}{l|}{0.37***} & \multicolumn{1}{l}{0.20***} & \multicolumn{1}{l}{$0.07^{\circ}$} & \multicolumn{1}{l|}{0.27***} & 0.18*** & 0.08* & 0.26*** & 0.17*** & 0.08* & 0.25*** \\
		&       & \multicolumn{1}{l}{[13.92]} & \multicolumn{1}{l}{[1.66]} & \multicolumn{1}{l|}{[6.19]} & \multicolumn{1}{l}{[9.66]} & \multicolumn{1}{l}{[1.75]} & \multicolumn{1}{l|}{[5.29]} & [8.76] & [2.12] & [5.24] & [8.05] & [2.10] & [4.89] \\
		& \multirow{2}[0]{*}{L} & \multicolumn{1}{l}{0.89***} & \multicolumn{1}{l}{$0.25^{\circ}$} & \multicolumn{1}{l|}{1.14***} & \multicolumn{1}{l}{0.82***} & \multicolumn{1}{l}{$0.27^{\circ}$} & \multicolumn{1}{l|}{1.09***} & 0.84*** & 0.35* & 1.19*** & 0.83*** & 0.39* & 1.22*** \\
		&       & \multicolumn{1}{l}{[29.3]} & \multicolumn{1}{l}{[1.71]} & \multicolumn{1}{l|}{[7.40]} & \multicolumn{1}{l}{[26.9]} & \multicolumn{1}{l}{[1.81]} & \multicolumn{1}{l|}{[6.51]} & [27.4] & [2.33] & [7.4] & [26.33] & [2.26] & [6.6] \\
		& \multirow{2}[0]{*}{G} & \multicolumn{1}{l}{} & \multicolumn{1}{l}{} & \multicolumn{1}{l|}{} & \multicolumn{1}{l}{0.17***} & \multicolumn{1}{l}{$0.06^{\circ}$} & \multicolumn{1}{l|}{0.23***} &       &       &       & 0.05* & 0.02  & 0.08* \\
		&       &       &       &       & \multicolumn{1}{l}{[8.49]} & \multicolumn{1}{l}{[1.83]} & \multicolumn{1}{l|}{[5.48]} &       &       &       & [2.17] & [1.57] & [2.09] \\
		& \multirow{2}[1]{*}{S} & \multicolumn{1}{l}{} & \multicolumn{1}{l}{} & \multicolumn{1}{l|}{} & \multicolumn{1}{l}{} & \multicolumn{1}{l}{} & \multicolumn{1}{l|}{} & 0.17*** & 0.07* & 0.24*** & 0.15*** & 0.07* & 0.22*** \\
		&       &       &       &       &       &       &       & [10.55] & [2.29] & [6.3] & [7.43] & [2.17] & [5.02] \\
		\hline
		\multirow{8}[2]{*}{Cut-Off = 50 km} & \multirow{2}[1]{*}{K} & \multicolumn{1}{l}{0.29***} & \multicolumn{1}{l}{$0.12^{\circ}$} & \multicolumn{1}{l|}{0.41***} & \multicolumn{1}{l}{0.21***} & \multicolumn{1}{l}{$0.09^{\circ}$} & \multicolumn{1}{l|}{0.30***} & 0.19*** & 0.11* & 0.29*** & 0.18*** & 0.11* & 0.29*** \\
		&       & \multicolumn{1}{l}{[14.31]} & \multicolumn{1}{l}{[1.77]} & \multicolumn{1}{l|}{[5.23]} & \multicolumn{1}{l}{[9.66]} & \multicolumn{1}{l}{[1.72]} & \multicolumn{1}{l|}{[4.44]} & [8.79] & [2.02] & [4.46] & [8.7] & [2.17] & [4.49] \\
		& \multirow{2}[0]{*}{L} & \multicolumn{1}{l}{0.88***} & \multicolumn{1}{l}{$0.35^{\circ}$} & \multicolumn{1}{l|}{1.24***} & \multicolumn{1}{l}{0.81***} & \multicolumn{1}{l}{$0.36^{\circ}$} & \multicolumn{1}{l|}{1.17***} & 0.83*** & 0.48* & 1.31*** & 0.82*** & 0.53* & 1.35*** \\
		&       & \multicolumn{1}{l}{[30.17]} & \multicolumn{1}{l}{[1.80]} & \multicolumn{1}{l|}{[5.78]} & \multicolumn{1}{l}{[26.02]} & \multicolumn{1}{l}{[1.78]} & \multicolumn{1}{l|}{[5.16]} & [27.39] & [2.2] & [5.72] & [26.19] & [2.37] & [5.69] \\
		& \multirow{2}[0]{*}{G} & \multicolumn{1}{l}{} & \multicolumn{1}{l}{} & \multicolumn{1}{l|}{} & \multicolumn{1}{l}{0.17***} & \multicolumn{1}{l}{$0.08^{\circ}$} & \multicolumn{1}{l|}{0.25***} &       &       &       & 0.05* & 0.03  & $0.08^{\circ}$ \\
		&       &       &       &       & \multicolumn{1}{l}{[7.70]} & \multicolumn{1}{l}{[1.70]} & \multicolumn{1}{l|}{[4.20]} &       &       &       & [2.11] & [1.51] & [1.96] \\
		& \multirow{2}[1]{*}{S} & \multicolumn{1}{l}{} & \multicolumn{1}{l}{} & \multicolumn{1}{l|}{} & \multicolumn{1}{l}{} & \multicolumn{1}{l}{} & \multicolumn{1}{l|}{} & 0.17*** & 0.10* & 0.27*** & 0.15*** & 0.10* & 0.24*** \\
		&       &       &       &       &       &       &       & [11.05] & [2.14] & [5.1] & [7.5] & [2.23] & [4.49] \\
		\hline
	\end{tabular}%
}
		\label{tab:marg}%
	\caption*{\footnotesize {\textbf{Note:} Table \ref{tab:marg} presents the results of the decomposition in direct (D), indirect (I) and total effects (T). Estimates are considered in terms of elasticities, while z-values are in square brackets. The z-values and p-values are estimated by Bootstrap. Statistical significance: *** <0.001, ** 0.01, * 0.05, $^{\circ}$ 0.1 }}
\end{table}%

\begin{table}[htbp]
\centering
\caption{Spatial Quantile Regression}
\noindent\resizebox{\linewidth}{!}{
	\begin{tabular}{|l|c|rrr|rrr|rrr|}
		\cline{3-11}    \multicolumn{1}{r}{} &       & \multicolumn{3}{c|}{Cut-off = 33 Km} & \multicolumn{3}{c|}{Cut-off = 41 Km} & \multicolumn{3}{c|}{Cut-off = 50 Km} \\
		\cline{2-11}    \multicolumn{1}{r|}{} & \multicolumn{1}{c|}{Q} & \multicolumn{1}{l}{Coeff} & \multicolumn{1}{l}{Z-val.} & \multicolumn{1}{l|}{P-val.} & \multicolumn{1}{l}{Coeff} & \multicolumn{1}{l}{Z-val.} & \multicolumn{1}{l|}{P-val.} & \multicolumn{1}{l}{Coeff} & \multicolumn{1}{l}{Z-val.} & \multicolumn{1}{l|}{P-val.} \\
		\hline
		K     & \multirow{5}[2]{*}{0.10} & 0.16  & 3.82  & 0.00  & 0.17  & 3.95  & 0.00  & 0.19  & 4.24  & 0.00 \\
		L     &       & 0.80  & 13.65 & 0.00  & 0.79  & 13.25 & 0.00  & 0.78  & 12.86 & 0.00 \\
		G     &       & -0.05 & -0.89 & 0.37  & -0.07 & -1.20 & 0.23  & -0.06 & -1.03 & 0.31 \\
		S     &       & 0.31  & 6.61  & 0.00  & 0.33  & 6.91  & 0.00  & 0.33  & 6.83  & 0.00 \\
		$\rho$   &       & 0.28  & 2.26  & 0.02  & 0.28  & 1.91  & 0.06  & 0.35  & 2.28  & 0.02 \\
		\hline
		K     & \multirow{5}[2]{*}{0.25} & 0.13  & 4.56  & 0.00  & 0.12  & 3.99  & 0.00  & 0.13  & 4.37  & 0.00 \\
		L     &       & 0.86  & 22.96 & 0.00  & 0.87  & 22.43 & 0.00  & 0.83  & 20.59 & 0.00 \\
		G     &       & 0.01  & 0.19  & 0.85  & 0.02  & 0.48  & 0.63  & 0.02  & 0.79  & 0.43 \\
		S     &       & 0.20  & 6.04  & 0.00  & 0.20  & 6.42  & 0.00  & 0.20  & 6.10  & 0.00 \\
		$\rho$   &       & 0.23  & 2.49  & 0.01  & 0.26  & 2.32  & 0.02  & 0.52  & 3.96  & 0.00 \\
		\hline
		K     & \multirow{5}[2]{*}{0.50} & 0.15  & 7.19  & 0.00  & 0.15  & 6.80  & 0.00  & 0.15  & 7.03  & 0.00 \\
		L     &       & 0.80  & 23.98 & 0.00  & 0.82  & 26.32 & 0.00  & 0.81  & 25.90 & 0.00 \\
		G     &       & 0.05  & 2.55  & 0.01  & 0.04  & 1.77  & 0.08  & 0.05  & 2.55  & 0.01 \\
		S     &       & 0.14  & 7.79  & 0.00  & 0.15  & 8.14  & 0.00  & 0.15  & 7.68  & 0.00 \\
		$\rho$   &       & 0.26  & 3.78  & 0.00  & 0.28  & 3.09  & 0.00  & 0.42  & 3.82  & 0.00 \\
		\hline
		K     & \multirow{5}[2]{*}{0.75} & 0.16  & 7.43  & 0.00  & 0.16  & 6.73  & 0.00  & 0.16  & 6.70  & 0.00 \\
		L     &       & 0.81  & 25.48 & 0.00  & 0.81  & 26.52 & 0.00  & 0.80  & 24.15 & 0.00 \\
		G     &       & 0.06  & 1.72  & 0.09  & 0.06  & 1.75  & 0.08  & 0.05  & 1.32  & 0.19 \\
		S     &       & 0.13  & 5.56  & 0.00  & 0.13  & 5.34  & 0.00  & 0.13  & 5.96  & 0.00 \\
		$\rho$   &       & 0.30  & 3.73  & 0.00  & 0.36  & 3.88  & 0.00  & 0.36  & 3.62  & 0.00 \\
		\hline
		K     & \multirow{5}[2]{*}{0.90} & 0.17  & 5.55  & 0.00  & 0.16  & 5.91  & 0.00  & 0.16  & 5.63  & 0.00 \\
		L     &       & 0.81  & 16.88 & 0.00  & 0.85  & 17.70 & 0.00  & 0.83  & 16.69 & 0.00 \\
		G     &       & 0.10  & 2.95  & 0.00  & 0.09  & 2.73  & 0.01  & 0.11  & 3.24  & 0.00 \\
		S     &       & 0.10  & 3.66  & 0.00  & 0.09  & 3.53  & 0.00  & 0.10  & 4.30  & 0.00 \\
		$\rho$   &       & 0.31  & 2.98  & 0.00  & 0.25  & 1.97  & 0.05  & 0.40  & 3.10  & 0.00 \\
		\hline
	\end{tabular}
}
	\label{tab:sqr}
	\caption*{\footnotesize {\textbf{Note:} Table \ref{tab:sqr} presents the results of Spatial AutoRegressive Quantile Regression for 2008. Estimates are considered in terms of elasticities. Three distinct cutt-off are considered: 33 km corresponding to the minimum distance for which every unit has at least one neighbour, 41 km (1.25* min. dist) and 50 km (1.5* min dist.). In Table \ref{tab:sqr} Q indicates the quantiles and are reported only the estimates for the extreme of the distribution (0.1 and 0.9) and the quartiles (0.25;0.5;0.75).}}
\end{table}%

\begin{table}[htbp!]
	\centering
	\caption{Spatial Quantile Regression - Marginal Impact}
	\noindent\resizebox{\linewidth}{!}{
		\begin{tabular}{|c|llll|llll|llll|c|}
		\cline{2-13}    \multicolumn{1}{r|}{} & \multicolumn{4}{c|}{Distance = 33 km} & \multicolumn{4}{c|}{Distance = 41 km} & \multicolumn{4}{c|}{Distance = 50 km} & \multicolumn{1}{r}{} \\
		\cline{2-14}    \multicolumn{1}{r|}{} & \multicolumn{1}{c}{K} & \multicolumn{1}{c}{L} & \multicolumn{1}{c}{G} & \multicolumn{1}{c|}{S} & \multicolumn{1}{c}{K} & \multicolumn{1}{c}{L} & \multicolumn{1}{c}{G} & \multicolumn{1}{c|}{S} &\multicolumn{1}{c}{K} & \multicolumn{1}{c}{L} & \multicolumn{1}{c}{G} & \multicolumn{1}{c|}{S} & Q \\
		\hline
		\multirow{2}[0]{*}{D} & 0.16*** & 0.80*** & -0.05  & 0.31*** & 0.17*** & 0.80*** & -0.07  & 0.33*** & 0.19*** & 0.78*** & -0.06  & 0.33*** & \multirow{6}[1]{*}{0.1} \\
		& [3.33] & [15.11] & [-1.00] & [7.55] & [4.03] & [13.5] & [-1.26] & [7.37] & [4.61] & [14.64] & [-1.09] & [6.49] &  \\
		\multirow{2}[0]{*}{I} & 0.06   & 0.31   & -0.02  & 0.12   & 0.07   & 0.31   & -0.03  & 0.13   & 0.1    & 0.42   & -0.03  & 0.18   &  \\
		& [1.56] & [1.64] & [-0.84] & [1.60] & [1.36] & [1.45] & [-0.86] & [1.37] & [1.09] & [1.29] & [-0.70] & [1.18] &  \\
		\multirow{2}[1]{*}{T} & 0.22** & 1.12*** & -0.07  & 0.44*** & 0.24*** & 1.11*** & -0.10  & 0.46*** & 0.29*  & 1.21*** & -0.09  & 0.50** &  \\
		& [3.05] & [5.22] & [-1.00] & [4.45] & [3.06] & [4.96] & [-1.21] & [3.82] & [2.29] & [3.25] & [-0.99] & [2.64] &  \\
		\hline
		\multirow{2}[1]{*}{D} & 0.13*** & 0.87*** & 0.01   & 0.20*** & 0.12*** & 0.87*** & 0.02   & 0.20*** & 0.13*** & 0.84*** & 0.02   & 0.20*** & \multirow{6}[2]{*}{0.25} \\
		& [4.13] & [22.93] & [0.16] & [6.41] & [3.94] & [24.74] & [0.38] & [6.16] & [4.55] & [19.75] & [0.82] & [7.00] &  \\
		\multirow{2}[0]{*}{I} & 0.04   & 0.25$^{\circ}$  & 0      & 0.06$^{\circ}$  & 0.04   & 0.30$^{\circ}$  & 0.01   & 0.07$^{\circ}$  & 0.14   & 0.9    & 0.02   & 0.21   &  \\
		& [1.50] & [1.87] & [0.15] & [1.81] & [1.49] & [1.7]  & [0.31] & [1.7]  & [1.02] & [1.35] & [0.65] & [1.26] &  \\
		\multirow{2}[1]{*}{T} & 0.17** & 1.12*** & 0.01   & 0.26*** & 0.16*** & 1.17*** & 0.02   & 0.27*** & 0.28   & 1.74*  & 0.05   & 0.41*  &  \\
		& [3.29] & [7.98] & [0.16] & [5.12] & [3.18] & [6.37] & [0.37] & [4.82] & [1.62] & [2.39] & [0.75] & [2.12] &  \\
		\hline
		\multirow{2}[1]{*}{D} & 0.15*** & 0.81*** & 0.05$^{\circ}$  & 0.14*** & 0.15*** & 0.82*** & 0.04$^{\circ}$  & 0.15*** & 0.15*** & 0.81*** & 0.05*  & 0.15*** & \multirow{6}[2]{*}{0.5} \\
		& [7.07] & [24.41] & [1.93] & [6.56] & [6.95] & [27.63] & [1.79] & [7.53] & [6.59] & [23.73] & [2.44] & [7.23] &  \\
		\multirow{2}[0]{*}{I} & 0.05*  & 0.28** & 0.02   & 0.05** & 0.06$^{\circ}$  & 0.32*  & 0.01   & 0.06*  & 0.11$^{\circ}$  & 0.58*  & 0.04   & 0.11*  &  \\
		& [2.26] & [2.50] & [1.41] & [2.69] & [1.89] & [2.15] & [1.24] & [2.27] & [1.77] & [2.23] & [1.34] & [2.26] &  \\
		\multirow{2}[1]{*}{T} & 0.21*** & 1.08*** & 0.06$^{\circ}$  & 0.19*** & 0.20*** & 1.15*** & 0.05$^{\circ}$  & 0.21*** & 0.25** & 1.39*** & 0.08$^{\circ}$  & 0.25*** &  \\
		& [5.43] & [9.46] & [1.85] & [6.85] & [4.54] & [7.4]  & [1.67] & [6.73] & [3.24] & [5.06] & [1.88] & [4.78] &  \\
		\hline
		\multirow{2}[1]{*}{D} & 0.16*** & 0.81*** & 0.06   & 0.13*** & 0.16*** & 0.81*** & 0.06$^{\circ}$  & 0.13*** & 0.16*** & 0.80*** & 0.05   & 0.13*** & \multirow{6}[2]{*}{0.75} \\
		& [6.28] & [25.45] & [1.60] & [5.40] & [6.51] & [25.37] & [1.81] & [5.5]  & [6.22] & [27.67] & [1.28] & [5.25] &  \\
		\multirow{2}[0]{*}{I} & 0.07*  & 0.34** & 0.02   & 0.06*  & 0.09*  & 0.44** & 0.03   & 0.07*  & 0.09*  & 0.44** & 0.03   & 0.07*  &  \\
		& [2.55] & [3.01] & [1.30] & [2.47] & [2.12] & [2.61] & [1.41] & [2.23] & [2.11] & [2.57] & [1.08] & [2.29] &  \\
		\multirow{2}[1]{*}{T} & 0.22*** & 1.15*** & 0.08   & 0.19*** & 0.25*** & 1.25*** & 0.09$^{\circ}$  & 0.19*** & 0.25*** & 1.24*** & 0.07   & 0.21*** &  \\
		& [5.09] & [9.44] & [1.54] & [4.46] & [4.12] & [7.59] & [1.75] & [4.23] & [3.98] & [6.86] & [1.25] & [4.13] &  \\
		\hline
		\multirow{2}[1]{*}{D} & 0.17*** & 0.81*** & 0.10** & 0.10*** & 0.16*** & 0.85*** & 0.09** & 0.09*** & 0.16*** & 0.83*** & 0.11** & 0.10*** & \multirow{6}[2]{*}{0.9} \\
		& [5.87] & [17.41] & [2.97] & [4.06] & [5.61] & [17.12] & [2.75] & [4.19] & [5.70] & [19.80] & [3.05] & [3.9]  &  \\
		\multirow{2}[0]{*}{I} & 0.07$^{\circ}$  & 0.36*  & 0.04$^{\circ}$  & 0.04$^{\circ}$  & 0.05   & 0.28   & 0.03   & 0.03   & 0.10$^{\circ}$  & 0.55*  & 0.07   & 0.06*  &  \\
		& [1.93] & [2.22] & [1.72] & [1.84] & [1.18] & [1.27] & [1.1]  & [1.27] & [1.89] & [2.07] & [1.40] & [1.94] &  \\
		\multirow{2}[1]{*}{T} & 0.24*** & 1.18*** & 0.14** & 0.14*** & 0.21*** & 1.13*** & 0.12*  & 0.12*** & 0.26*** & 1.38*** & 0.18*  & 0.16** &  \\
		& [4.13] & [7.01] & [2.75] & [3.43] & [3.26] & [4.51] & [2.31] & [3.23] & [3.63] & [5.04] & [2.19] & [3.35] &  \\
		\hline
	\end{tabular}
}
			\label{tab:marg_sqr}%
	\caption*{\footnotesize {\textbf{Note:} Table \ref{tab:marg_sqr} presents the results of the decomposition in direct (D), indirect (I) and total effects (T). Q indicated the considered quantile. Estimates are considered in terms of elasticities, while z-values are in square brackets. The z-values and p-values are estimated by Bootstrap.\\
    Statistical significance: *** <0.001, ** 0.01, * 0.05, $^{\circ}$ 0.1 }}
\end{table}%

\begin{table}[htbp]
	\centering
	\caption{Spatial Quantile Regression: Robustness Check}
	\noindent\resizebox{\linewidth}{!}{
	\begin{tabular}{|l|lll|lll|lll|c|}
		\cline{2-10}    \multicolumn{1}{r|}{} & \multicolumn{3}{c|}{Cut-off Distance = 33 km} & \multicolumn{3}{c|}{Cut-off Distance = 41 km} & \multicolumn{3}{c|}{Cut-off Distance = 50 km} & \multicolumn{1}{r}{} \\
		\cline{2-11}    \multicolumn{1}{r|}{} & Estimate & Std.Error & P-value & Estimate & Std.Error & P-value & Estimate & Std.Error & P-value & Q \\
		\hline
		K      & 0.181  & 0.047  & 0.000  & 0.174  & 0.054  & 0.001  & 0.166  & 0.057  & 0.003  & \multirow{5}[2]{*}{0.1} \\
		L      & 0.823  & 0.063  & 0.000  & 0.808  & 0.072  & 0.000  & 0.800  & 0.070  & 0.000  &  \\
		G      & -0.104 & 0.060  & 0.082  & -0.087 & 0.075  & 0.245  & -0.072 & 0.080  & 0.367  &  \\
		S      & 0.326  & 0.051  & 0.000  & 0.324  & 0.056  & 0.000  & 0.336  & 0.072  & 0.000  &  \\
		$\rho$     & 0.230  & 0.131  & 0.078  & 0.160  & 0.188  & 0.394  & 0.300  & 0.182  & 0.099  &  \\
		\hline
		K      & 0.122  & 0.031  & 0.000  & 0.126  & 0.030  & 0.000  & 0.122  & 0.028  & 0.000  & \multirow{5}[2]{*}{0.25} \\
		L      & 0.845  & 0.044  & 0.000  & 0.860  & 0.044  & 0.000  & 0.853  & 0.038  & 0.000  &  \\
		G      & 0.001  & 0.031  & 0.985  & 0.014  & 0.030  & 0.645  & 0.013  & 0.030  & 0.653  &  \\
		S      & 0.211  & 0.040  & 0.000  & 0.197  & 0.034  & 0.000  & 0.206  & 0.036  & 0.000  &  \\
		$\rho$     & 0.250  & 0.116  & 0.031  & 0.260  & 0.139  & 0.062  & 0.430  & 0.096  & 0.000  &  \\
		\hline
		K      & 0.149  & 0.021  & 0.000  & 0.147  & 0.022  & 0.000  & 0.153  & 0.022  & 0.000  & \multirow{5}[2]{*}{0.5} \\
		L      & 0.800  & 0.032  & 0.000  & 0.814  & 0.031  & 0.000  & 0.809  & 0.033  & 0.000  &  \\
		G      & 0.040  & 0.024  & 0.091  & 0.035  & 0.023  & 0.139  & 0.037  & 0.024  & 0.116  &  \\
		S      & 0.142  & 0.020  & 0.000  & 0.150  & 0.020  & 0.000  & 0.149  & 0.020  & 0.000  &  \\
		$\rho$     & 0.280  & 0.077  & 0.000  & 0.310  & 0.089  & 0.000  & 0.400  & 0.099  & 0.000  &  \\
		\hline
		K      & 0.151  & 0.023  & 0.000  & 0.153  & 0.024  & 0.000  & 0.157  & 0.024  & 0.000  & \multirow{5}[2]{*}{0.75} \\
		L      & 0.809  & 0.032  & 0.000  & 0.809  & 0.031  & 0.000  & 0.802  & 0.030  & 0.000  &  \\
		G      & 0.060  & 0.030  & 0.046  & 0.064  & 0.032  & 0.043  & 0.059  & 0.031  & 0.058  &  \\
		S      & 0.130  & 0.020  & 0.000  & 0.129  & 0.019  & 0.000  & 0.129  & 0.020  & 0.000  &  \\
		$\rho$     & 0.290  & 0.069  & 0.000  & 0.320  & 0.079  & 0.000  & 0.380  & 0.085  & 0.000  &  \\
		\hline
		K      & 0.172  & 0.020  & 0.000  & 0.162  & 0.021  & 0.000  & 0.157  & 0.021  & 0.000  & \multirow{5}[2]{*}{0.9} \\
		L      & 0.792  & 0.034  & 0.000  & 0.817  & 0.036  & 0.000  & 0.823  & 0.037  & 0.000  &  \\
		G      & 0.116  & 0.033  & 0.001  & 0.114  & 0.033  & 0.001  & 0.110  & 0.032  & 0.001  &  \\
		S      & 0.086  & 0.024  & 0.000  & 0.084  & 0.023  & 0.000  & 0.085  & 0.022  & 0.000  &  \\
		$\rho$     & 0.350  & 0.114  & 0.002  & 0.300  & 0.128  & 0.019  & 0.360  & 0.143  & 0.012  &  \\
		\hline
	\end{tabular}
}
	\label{tab:rob_sqr_CH}%
\end{table}%

\end{document}